\documentclass[12pt]{article}
\usepackage {amsmath}
\usepackage{amssymb}
\usepackage{graphicx}
\newcount\timecount
\newcount\hours \newcount\minutes  \newcount\temp \newcount\pmhours
\hours = \time
\divide\hours by 60
\temp = \hours
\multiply\temp by 60
\minutes = \time
\advance\minutes by -\temp
\def\hour{\the\hours}
\def\minute{\ifnum\minutes<10 0\the\minutes
            \else\the\minutes\fi}
\def\clock{
\ifnum\hours=0 12:\minute\ AM
\else\ifnum\hours<12 \hour:\minute\ AM
      \else\ifnum\hours=12 12:\minute\ PM
            \else\ifnum\hours>12
                 \pmhours=\hours
                 \advance\pmhours by -12
                 \the\pmhours:\minute\ PM
                 \fi
            \fi
      \fi
\fi
}

\def\monthname{\relax\ifcase\month 0/\or January\or February\or
   March\or April\or May\or June\or July\or August\or September\or
   October\or November\or December\else\number\month/\fi}

\def\bold#1{\setbox0=\hbox{$#1$}%
     \kern-.025em\copy0\kern-\wd0
     \kern.05em\copy0\kern-\wd0
     \kern-.025em\raise.0433em\box0 }

\def\beq{\begin{equation}}
\def\eeq{\end{equation}}
\def\beqn{\begin{eqnarray}}
\def\eeqn{\end{eqnarray}}

\def\ga{\mathrel{\raise.3ex\hbox{$>$\kern-.75em\lower1ex\hbox{$\sim$}}}}
\def\la{\mathrel{\raise.3ex\hbox{$<$\kern-.75em\lower1ex\hbox{$\sim$}}}}
\def\gev{{\rm \, Ge\kern-0.125em V}}
\def\tev{{\rm \, Te\kern-0.125em V}}
\def\gyr{{\rm \, G\kern-0.125em yr}}

\def\gappeq{\mathrel{\rlap {\raise.5ex\hbox{$>$}}
{\lower.5ex\hbox{$\sim$}}}}
\def\lappeq{\mathrel{\rlap{\raise.5ex\hbox{$<$}}
{\lower.5ex\hbox{$\sim$}}}}
\def\Toprel#1\over#2{\mathrel{\mathop{#2}\limits^{#1}}}

\def\avg#1{\left\langle #1 \right\rangle}

\def\m12{m_{1\!/2}}

\def\bea{\begin{eqnarray}}
\def\eea{\end{eqnarray}}

\begin{document}
\begin{titlepage}
\pagestyle{empty}
\baselineskip=21pt
\rightline{KCL-PH-TH/2015-43, LCTS/2015-32, CERN-PH-TH/2015-238}
\rightline{UMN--TH--3504/15, FTPI--MINN--15/43}
\vskip 0.2in
\begin{center}
{\large{\bf Scenarios for Gluino Coannihilation}}
\end{center}
\begin{center}
\vskip 0.15in
{\bf John~Ellis}$^{1,2}$,
{\bf Jason~L.~Evans}$^{3,4}$,
{\bf Feng~Luo}$^{2}$
and {\bf Keith~A.~Olive}$^{3,4}$
\vskip 0.2in
{\small {\it
$^1${Theoretical Particle Physics and Cosmology Group, Department of
  Physics, \\ King's College London, London WC2R 2LS, United Kingdom}\\
  \vspace{0.25cm}
$^2${Theory Division, CERN, CH-1211 Geneva 23, Switzerland}\\
  \vspace{0.25cm}
$^3${School of Physics and Astronomy, University of Minnesota, Minneapolis, MN 55455, USA}\\
  \vspace{0.25cm}
$^4${William I. Fine Theoretical Physics Institute, School of Physics and Astronomy,\\
  \vspace{-0.25cm}
University of Minnesota, Minneapolis, MN 55455, USA}
}}

\vskip 0.2in
{\bf Abstract}
\end{center}
\baselineskip=18pt \noindent

We study supersymmetric scenarios in which the gluino is the next-to-lightest
supersymmetric particle (NLSP), with a mass sufficiently close to that of the lightest
supersymmetric particle (LSP) that gluino coannihilation becomes important.
One of these scenarios is the MSSM with soft supersymmetry-breaking squark and
slepton masses that are universal at an input GUT renormalization scale, but with
non-universal gaugino masses. The other scenario is an extension of the MSSM
to include vector-like supermultiplets. In both scenarios, we identify the regions of
parameter space where gluino coannihilation is important, and discuss their relations
to other regions of parameter space where other mechanisms bring the dark
matter density into the range allowed by cosmology. In the case of the non-universal
MSSM scenario, we find that the allowed range of parameter space is constrained
by the requirement of electroweak symmetry breaking, the avoidance of a charged
LSP and the measured mass of the Higgs boson, in particular, as well as the appearance
of other dark matter (co)annihilation processes. Nevertheless, LSP masses $m_\chi \lesssim 8$~TeV
with the correct dark matter density are quite possible. In the case of pure gravity mediation with 
additional vector-like supermultiplets, changes to the anomaly-mediated gluino mass and the threshold effects
associated with these states can make the gluino almost degenerate with the LSP, and we find a similar upper bound.

\vskip 0.2in
\leftline{October 2015}
\end{titlepage}
\baselineskip=18pt

\section{Introduction}

The absence of supersymmetry so far, at LHC Run~I \cite{ATLAS20,CMS20} and elsewhere, raises the
question where, if anywhere, is it hiding. There are scenarios for which, at
least, some supersymmetric particles were produced in LHC Run~I, but have
been overlooked. Examples include models where $R$-parity is violated \cite{lhcr},
or the spectra are compressed~\cite{lhcc,compress}. Alternatively, sparticles might be too heavy to have
been detected at LHC Run~I, but might be within range of future LHC runs \cite{interplay}. It is
also possible that supersymmetric particles may lie beyond the reach of the LHC
altogether, and require a future higher-energy $pp$ collider for its detection.

If one assumes that $R$-parity is conserved, the lightest supersymmetric
particle (LSP) must be stable, and hence make at least a contribution to the
cosmological cold dark matter density \cite{ehnos}. The total density of cold dark matter
is very tightly constrained by measurements of the cosmic microwave
background radiation \cite{Planck15}. It is clear, therefore, that the parameters of generic models
are constrained in very specific ways in order to realize the correct dark matter density \cite{eo6,elos,mc12,other}.
Moreover, this parameter space with the correct density is likely to be found in a region of parameter space
where the density varies rapidly with the parameters. In these cases, regions
where the LSP contributes only a fraction of the cold dark matter density will
have parameters similar to those regions yielding the correct total density.

This sensitivity of the dark matter density to parameters are particularly relevant for models with compressed and/or very
heavy spectra that have survived LHC searches. Examples of
specific choices of heavy spectra that yield the correct cosmological dark matter density
include scenarios in which the LSP, $\chi$, would have annihilated with itself through a direct-channel
boson such as the heavier neutral Higgs bosons $A$ and $H$ \cite{funnel}. Alternatively, there
might be one or more heavier supersymmetric particles that are
nearly degenerate with the LSP, $\chi$, and would have coannihilated with it in the early Universe
\cite{gs}.
There are several examples of possible coannihilating sparticles, including the
lighter stau, or possibly some other slepton \cite{stau}, the lighter stop squark \cite{stop,eds,raza}, the lighter chargino
\cite{chaco,eds}
and the gluino \cite{glu,shafi,hari,evo,deSimone:2014pda,liantao,raza,ELO}.

In most cases, coannihilation with a sparticle having stronger interactions extends
the allowed mass range of the LSP. The possibility of gluino coannihilation is
therefore particularly interesting since it interacts strongly suggesting it can accommodate a heavier LSP than
is possible from coannihilation with a stau or slepton. In fact, it has been shown
that a dark matter density realized by an LSP coannihilating with the gluino could lie well beyond the reach of the LHC,
with a mass as heavy as $m_\chi \lesssim 8$~TeV~\cite{ELO}.

The possibility of gluino coannihilation does not arise in the minimal supersymmetric extension of
the Standard Model (MSSM) with the soft supersymmetry-breaking parameters
constrained to be universal at the input GUT scale (the CMSSM) \cite{funnel,cmssm,efgo,cmssmwmap,eo6,ehow+}, nor in related
models with non-universal Higgs masses \cite{nonu,efgo,nuhm2,nuhm1,eosknuhm}. However, as we discuss in this paper,
gluino coannihilation can become important in variants of the MSSM with non-universal
gaugino masses, and in variations of pure gravity mediation (PGM) with non-minimal matter content such as additional vector-like supermultiplets \cite{hari,evo}.

The layout of this paper is as follows. In Section~2 we set out the coupled set of
Boltzmann equations that we use to calculate the relic LSP density, discussing the
circumstances under which the analysis can be reduced to a single Boltzmann equation for a particular combination
of sparticle abundances \cite{ELO}. Then, in Section~3 we discuss various scenarios with
non-universal gaugino masses in which gluino coannihilation can become important,
delineating the corresponding strips in parameter space and comparing their extents
with the results of~\cite{ELO}. We find that this scenario is constrained by the
requirement of consistent electroweak symmetry breaking (EWSB), by the measurement of
$m_H$, and by avoidance of a stop or chargino LSP. We give examples showing that the correct dark matter density is possible with
LSP masses as large as $8$~TeV. Section~4 contains a similar analysis of PGM models with
vector-like supermultiplets, focusing on an example with a single extra pair of
$\mathbf{10}$ and $\mathbf{\overline{10}}$ representations of SU(5).
Because the anomaly-mediated contribution to the gluino mass is zero in this case, 
threshold effects due to these additional states generate almost the entire gluino mass. 
This suppresses the gluino mass relative to those of the other gauginos, leading to near-degeneracy between
the gluino and the LSP. Neutralino dark matter candidates with similarly large values of $m_\chi$ are again possible.
Finally, Section~5 summarizes our conclusions and
discusses the prospects for discovering supersymmetry in these gluino coannihilation
scenarios.

\section{Calculations of Gluino Coannihilation}

In this section, we present general formulae for calculating the dark matter thermal relic density,
and then specialize it to the case of the gluino coannihilation scenarios we consider in this paper,
taking into account the effects of gluino-gluino bound states.

We consider $N$ $R$-odd species in the thermal bath in the early Universe. We assume that
the rates for interconverting the LSP (which is labeled as the first species, with mass $m_1$)
and the first $l$ species ($1 \le l \le N$) are sufficiently  large, compared to the Hubble expansion rate,
that to a very good approximation the ratios of densities are equal to the equilibrium ratios:
$n_i / n_1 = n_i^{eq} / n_1^{eq}$ for ($i = 1,...,l$).

For any of the $N$ species, the evolution of its number density is governed by the Boltzmann equation
\beq
{d n_i \over d t} + 3 H(T) n_i = - \sum_{j=1}^N {\avg {\sigma v}}_{ij \rightarrow SM} \left( n_i n_j - n_i^{eq} n_j^{eq} \right) - \sum_{\substack{j=1 \\ j \neq i}}^N {\avg {\Gamma}}_{i \rightarrow j} \left(n_i - n_i^{eq} \frac{n_j} {n_j^{eq}} \right)  \, ,
\label{eq:dnidt}
\eeq
where ${\avg {\sigma v}}_{ij \rightarrow SM}$ is the product of the
thermally-averaged relative velocity with the total cross section for the channels of $i$ and $j$ (co)annihilating
into Standard Model particles, and ${\avg {\Gamma}}_{i \rightarrow j}$ is the sum of all the thermally-averaged
decay and conversion rates for decay and conversion processes for which there is
one particle $i$ in the initial state and one particle $j$ in the final state,
with all other particles involved in these processes being Standard Model particles.
The relations between the thermally-averaged forward and backward reactions are used
in (\ref{eq:dnidt}), and we assume for all the Standard Model particles involved that
$n_{SM} = n_{SM}^{eq}$, so that, for example,
\beq
{\avg {\Gamma}}_{i \rightarrow j} n_i^{eq} = {\avg {\Gamma}}_{j \rightarrow i} n_j^{eq} \, .
\eeq
Written in terms of the yields, $Y_i \equiv n_i / s$, (\ref{eq:dnidt}) becomes
\beq
{d Y_i \over d x} = - \frac{x s}{H(m_1)} \left(1+\frac{T}{3 g_{\ast s}} \frac{d g_{\ast s}}{d T} \right)\left[\sum_{j=1}^N {\avg {\sigma v}}_{ij \rightarrow SM} \left( Y_i Y_j - Y_i^{eq} Y_j^{eq} \right) + \sum_{\substack{j=1 \\ j \neq i}}^N \frac{{\avg {\Gamma}}_{i \rightarrow j}}{s} \left(Y_i - Y_i^{eq} \frac{Y_j} {Y_j^{eq}} \right) \right]  \, ,
\label{eq:dyidt}
\eeq
where
\beq
x \equiv \frac{m_1}{T}, \, s = {2 \pi^2 \over 45} g_{\ast s} T^3, \; H(m_1) \equiv H(T) x^2 = ({4 \pi^3 G_N g_\ast \over 45})^{1 \over 2} m_1^2 \, ,
\eeq
and $g_{\ast s}$ and $g_\ast$ are the total numbers of effectively massless degrees of freedom associated with
the entropy density and the energy density, respectively.

Defining $\tilde{Y} \equiv \sum\limits_{i=1}^l Y_i$, $\Delta_i \equiv (m_i - m_1)/m_1$
and $\tilde{g}_{eff} \equiv \sum\limits_{i=1}^l g_i (1+\Delta_i)^{3/2} e^{-\Delta_i x}$, we have
\beq
\frac{Y_i^{eq}}{\tilde{Y}^{eq}} = \frac{g_i \left( \frac{m_i T}{2 \pi} \right)^{3/2}  e^{-m_i/T}}{\sum\limits_{j=1}^l g_j \left( \frac{m_j T}{2 \pi} \right)^{3/2} e^{-m_j/T}}
= \frac{g_i \left(1+\Delta_i \right)^{3/2} e^{-\Delta_i x}}{\tilde{g}_{eff}} \, .
\eeq
Using $Y_i/\tilde{Y} = Y_i^{eq}/\tilde{Y}^{eq}$ for $i = 1,...,l$ and summing over (\ref{eq:dyidt}) for the first $l$ species,
we find
\beqn
{d \tilde{Y} \over d x} &=& - \frac{x s}{H(m_1)} \left(1+\frac{T}{3 g_{\ast s}} \frac{d g_{\ast s}}{d T} \right)
\Bigg\{ {\avg {\tilde{\sigma}_{eff} v}} \left( \tilde{Y}^2  - \tilde{Y}_{eq}^2 \right)
 \nonumber  \\ &&
 + \sum_{k = l+1}^N \left[{\avg {{\sigma_k}_{eff} v}} \left( \tilde{Y} Y_k - \tilde{Y}^{eq} Y_k^{eq}\right)
 +  \frac{{\avg {\Gamma_k}_{eff}}}{s} \left(\tilde{Y} - \tilde{Y}^{eq} \frac{Y_k}{Y_k^{eq}} \right) \right] \Bigg\} \, ,
\label{eq:dytildedx}
\eeqn
while for each of the species $k$ ($l < k \le N$), we get
\beqn
{d Y_k \over d x} &=& - \frac{x s}{H(m_1)} \left(1+\frac{T}{3 g_{\ast s}} \frac{d g_{\ast s}}{d T} \right)
\Bigg\{ {\avg {{\sigma_k}_{eff} v}}  \left( \tilde{Y} Y_k  - \tilde{Y}^{eq} Y_k^{eq} \right)
 \nonumber  \\ &&
 + \sum_{j=l+1}^N {\avg {\sigma v}}_{kj \rightarrow SM} \left( Y_k Y_j - Y_k^{eq} Y_j^{eq} \right) -  \frac{{\avg {\Gamma_k}_{eff}}}{s}  \left(\tilde{Y} - \tilde{Y}^{eq} \frac{Y_k}{Y_k^{eq}} \right)
 \nonumber  \\ &&
 + \sum_{\substack{j=l+1 \\ j \neq k}}^N \frac{{\avg {\Gamma}}_{k \rightarrow j}}{s} \left(Y_k - Y_k^{eq} \frac{Y_j}{Y_j^{eq}} \right) \Bigg\} \, ,
\label{eq:dykdx}
\eeqn
where
\beqn
{\avg {{\sigma_k}_{eff} v}} &\equiv& \sum_{i=1}^l {\avg {\sigma v}}_{ik \rightarrow SM}  \frac{Y_i^{eq}}{\tilde{Y}^{eq}} = \sum_{i=1}^l {\avg {\sigma v}}_{ik \rightarrow SM}  \frac{g_i \left(1+\Delta_i \right)^{3/2} e^{-\Delta_i x}}{\tilde{g}_{eff}} \, ,
\\
{\avg {\Gamma_k}_{eff}} &\equiv& \sum_{i=1}^l {\avg \Gamma_{i \rightarrow k}} \frac{Y_i^{eq}}{\tilde{Y}^{eq}} = \sum_{i=1}^l {\avg \Gamma_{i \rightarrow k}} \frac{g_i \left(1+\Delta_i \right)^{3/2} e^{-\Delta_i x}}{\tilde{g}_{eff}} \, ,
\\
{\avg {\tilde{\sigma}_{eff} v}} &\equiv& \sum_{i,j=1}^l {\avg {\sigma v}}_{ij \rightarrow SM} \frac{Y_i^{eq} Y_j^{eq}}{\tilde{Y}_{eq}^2} = \sum_{i,j=1}^l {\avg {\sigma v}}_{ij \rightarrow SM} \frac{g_i g_j \left(1+\Delta_i \right)^{3/2} \left(1+\Delta_j \right)^{3/2} e^{-(\Delta_i + \Delta_j) x}}{\tilde{g}_{eff}^2} \, .
\nonumber \\
\eeqn
We note that in the case $N = l + 1$, the final term in (\ref{eq:dykdx}) does not appear.
In the case $N = l$, (\ref{eq:dytildedx}) does not have the two terms in the squared bracket,
and reverts to the familiar form for coannihilations when all the $N$ species are sufficiently coupled to the LSP.

We now specialize the above general formulae to the gluino coannihilation scenarios we consider in this paper.
First of all, following the discussion in~\cite{ELO}, the effect of gluino-gluino bound states on the calculation
of the dark matter relic density can be taken into account simply by modifying the Boltzmann equation
by including the Sommerfeld-enhanced thermal-averaged velocity-weighted gluino pair annihilation
cross section \cite{deSimone:2014pda}, which includes gluino-pair annihilation to two gluons and to all the quark anti-quark pair channels:
\beq
{\avg {\sigma v}}_{{\tilde g} {\tilde g} \rightarrow g g, q \bar{q}}
\rightarrow {\avg {\sigma v}}_{{\tilde g} {\tilde g} \, incl. \, \tilde{R}} \equiv {\avg {\sigma v}}_{{\tilde g} {\tilde g} \rightarrow g g, q \bar{q}} + {\avg {\sigma v}}_{bsf} {{\avg \Gamma}_{\tilde R}   \over {\avg \Gamma}_{\tilde R} + {\avg \Gamma}_{dis}}  \, ,
\label{eq:effglgl}
\eeq
where ${\avg {\sigma v}}_{bsf}$, ${\avg \Gamma}_{\tilde R}$ and ${\avg \Gamma}_{dis}$ are the thermally-averaged
formation cross section, decay rate and dissociation rate for the bound state ${\tilde R}$, respectively. The details of these quantities and the derivation of eq.~(\ref{eq:effglgl}) can be found in Section 3, 5 and Appendix B of~\cite{ELO}.

When the rate for interconverting the neutralino LSP and the gluino is sufficiently large, compared to the Hubble rate,
so that to a good approximation the relation $Y_{\tilde{g}} (T)/Y_1 (T) = Y_{\tilde{g}}^{eq} (T)/Y_1^{eq} (T)$
holds at all temperatures during which the sum of $Y_{\tilde{g}} (T)$ and $Y_1 (T)$ changes non-negligibly,
we can use a single Boltzmann equation to solve for the dark matter relic abundance,
including the gluino species in $\tilde{Y}$ and ${\avg {\tilde{\sigma}_{eff} v}}$ and using
(\ref{eq:dytildedx}) without the two terms in the squared bracket. Otherwise, one should use a
coupled set of Boltzmann equations, namely (\ref{eq:dytildedx}) and (\ref{eq:dykdx}), to solve for the dark matter
relic abundance. For the scenarios considered in this paper, any of the $R$-odd species
apart from the gluino is either sufficiently coupled to the LSP by having a Standard Model particle in the
propagator of a tree-level Feynman diagram describing its interconversion with the LSP,
or is so heavy compared to the LSP that it is effectively not participating coannihilations.
Therefore, when using a coupled set of Boltzmann equations, we have $N = l + 1$ in
(\ref{eq:dytildedx}) and (\ref{eq:dykdx}), and the index $k$ is for the gluino.

We end this Section by emphasizing that, in principle, the coupled set of Boltzmann equations
can always be used to solve for the dark matter
relic density, whether the rate for interconverting the gluino and the LSP is sufficiently large
compared to the Hubble rate or not. However, for the former case, solving a single Boltzmann equation is
usually easier than solving the coupled ones and requires less computing time.

\section{The Non-Universal MSSM Scenario}

It was assumed in~\cite{ELO} that the squarks were all degenerate with a common
mass $m_{\tilde q}$, and the effects of sparticles with only electroweak interactions
were neglected. It was found in~\cite{ELO} that in the presence of gluino coannihilation,
a Bino LSP, $\chi$, could be the dark matter of the universe if it weighed $\lesssim 8$~TeV, the exact value depending on the ratio
$m_{\tilde q}/m_\chi$, with smaller values of $m_\chi$ being found for $m_{\tilde q}/m_\chi
\lesssim 5$ and $\gtrsim 100$. Here we make a more complete study in a variant of the
MSSM with universal soft supersymmetry-breaking scalar masses $m_0$ and trilinear
couplings $A_0$, allowing a restricted form of non-universality in the gaugino sector
with $M_1 = M_2 \ne M_3$ at the input GUT scale. The results therefore depend on
$M_1/M_3$ as well as the usual CMSSM parameters $m_0, A_0$ and $\tan \beta$ (the ratio of MSSM Higgs vev's).
This is therefore a one-parameter extension of the CMSSM (with the new free parameter
being $M_3$) as is the NUHM1 (with the soft Higgs masses $m_1 = m_2 \ne m_0$)~\cite{nuhm1,eosknuhm}.
We consider in this section various $(M_1, M_3)$ planes for various choices of the
other parameters which illustrate the range of possibilities.

\begin{figure}
\begin{center}
\hspace{-0.6cm}
\includegraphics[height=7cm]{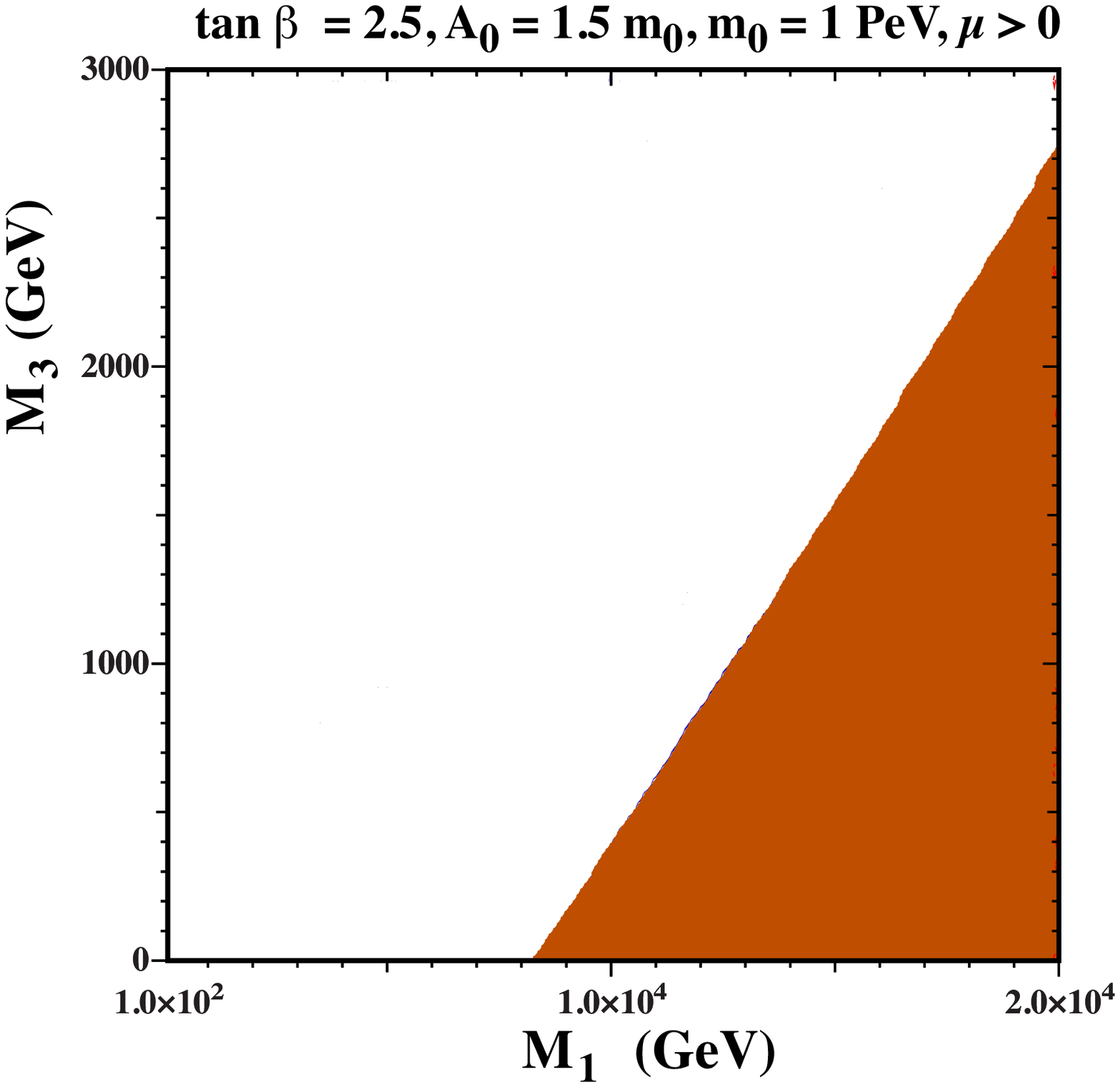}
\includegraphics[height=7cm]{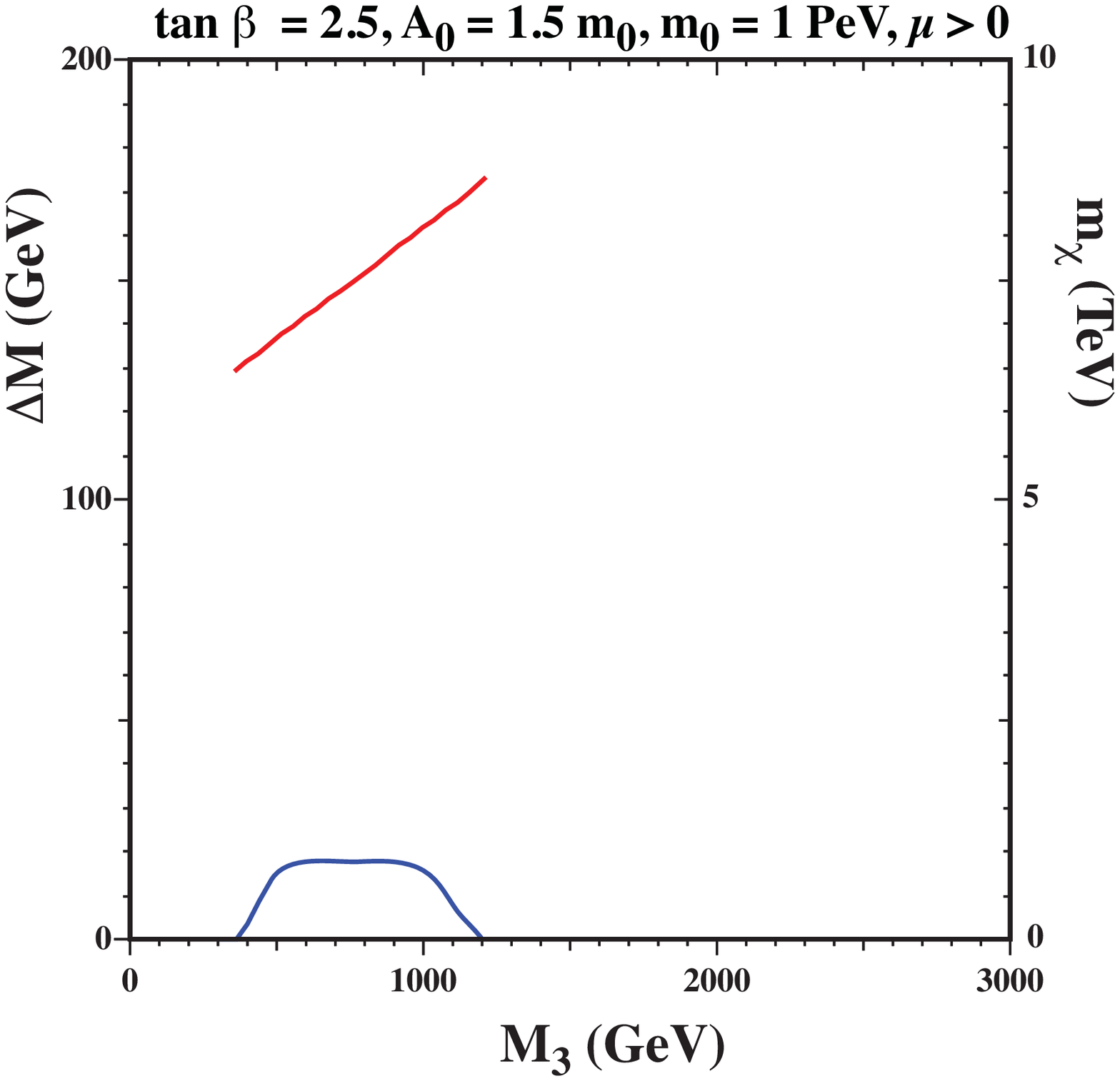}
\end{center}
\caption{\label{fig:1000}\it
The $(M_1, M_3)$ plane (left) for $m_0 = 1000$~TeV, $A_0/m_0 = 1.5$ and $\tan \beta = 2.5$.
The dark blue strip in the left panel shows where the relic
LSP density $\Omega_\chi h^2$ falls within the $\pm 3$-$\sigma$ range allowed by Planck
and other data, and the lightest neutralino is no longer the LSP in the regions shaded brick-red.
The right panel shows the gluino-neutralino mass difference (left axis, blue line) and the neutralino mass (right
axis, red line) as functions of $M_3$.}
\end{figure}

We first consider the example with $m_0 = 1000$~TeV, $A_0/m_0 = 1.5$ and $\tan \beta = 2.5$
shown in Fig.~\ref{fig:1000}. In the left panel and in subsequent figures, the regions where the relic
LSP density $\Omega_\chi h^2$ falls within the range allowed by Planck
and other data are shown as dark blue strips, and the regions where the lightest neutralino is
no longer the LSP are shaded brick-red. In this case, the gluino is the LSP in the shaded region.
Because of the scale of the plot, it is difficult to discern the relic density strip,
which lies very close to
the boundary of this region. However, we note that it lies to the left of the red shaded region only when $M_3$ is
between $\sim 400$ and $\sim 1200$~GeV, as shown in the right panel of Fig.~\ref{fig:1000} by the left axis and blue curve, which
shows the mass difference $\Delta M \equiv m_{\tilde g} - m_\chi$ between the gluino and the neutralino
along the coannihilation strip as
a function of the input gluino mass.  Also shown in the right panel (as a red line) is the neutralino mass as a function of $M_3$.
As seen in Fig. 6 in~\cite{ELO}, the choice of $m_0$ in Fig.~\ref{fig:1000} corresponds to values of $m_{\tilde q}/m_\chi$ extending
from beyond the plateau at small $M_3$ to values along the plateau at large $M_3$.
The gluino coannihilation strip therefore has two end-points where $\Delta M \to 0$,
corresponding to the limiting values $m_\chi \sim 6$~TeV and $m_\chi \sim 8$~TeV seen in the right panel of Fig.~\ref{fig:1000}:
for larger and smaller $M_3$, $\Delta M < 0$ and
the gluino is the LSP.

We note that the Higgs mass is relatively insensitive to the choice of $M_1$ and $M_3$, and therefore varies
very little across the plane with $\tan \beta, A_0$, and $m_0$ fixed~\footnote{We calculate the Higgs mass using the procedure outlined in \cite{Giudice:2011cg}.}. For the case shown in
Fig.~\ref{fig:1000}, we calculate $m_H \approx 126.3$ GeV, which is compatible with the
experimental measurement, within the theoretical uncertainties.
We do not show any other $(M_1, M_3)$ planes for $m_0 = 1000$~TeV, since the
possibilities are quite limited: there are no consistent solutions of the electroweak
symmetry-breaking conditions for much smaller values of $A_0/m_0 \lesssim 1$ and/or
larger values of $\tan \beta$, and $m_H$ is too large for larger values of $\tan \beta$  and/or $A_0/m_0$
(though it increases quite slowly with $A_0$).

\begin{figure}
\begin{center}
\begin{tabular}{c c}
\hspace{-0.6cm}
\includegraphics[height=7cm]{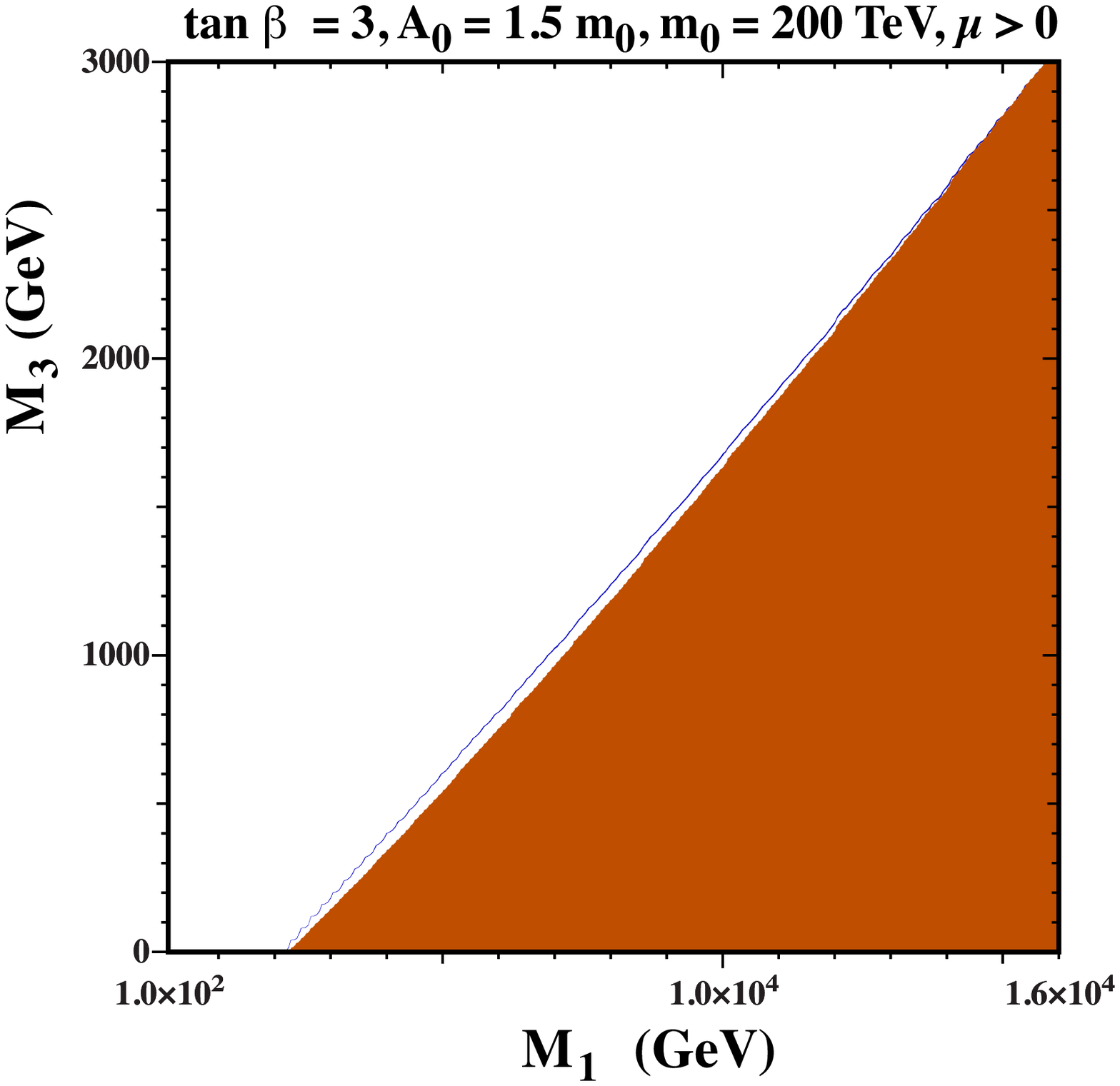} &
\hspace{-0.6cm}
\includegraphics[height=7cm]{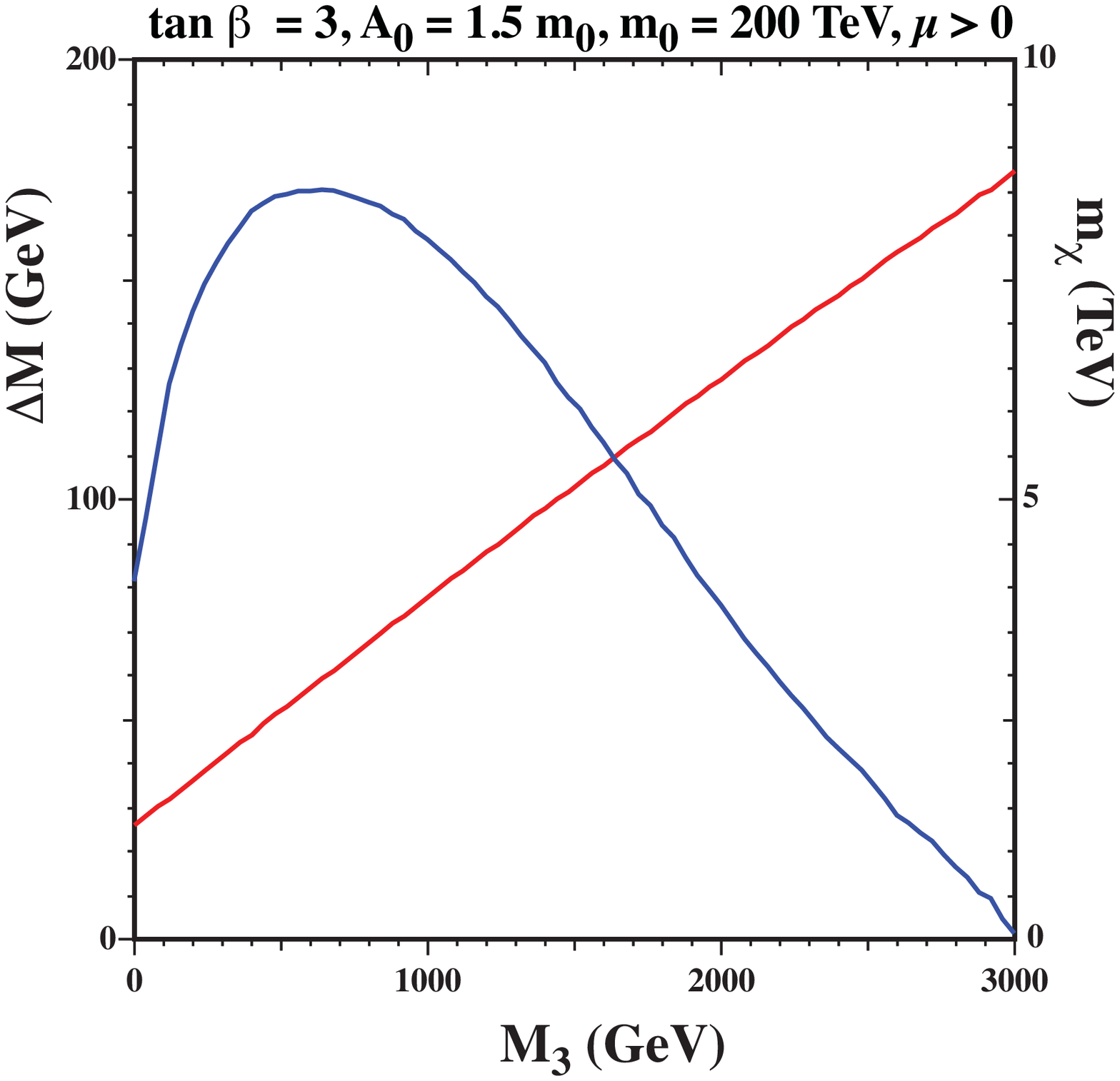} \\
\end{tabular}
\end{center}
\caption{\label{fig:200}\it
As for Fig.~\protect\ref{fig:1000}, but for $m_0 = 200$~TeV and $A_0/m_0 = 1.5$, with
$\tan \beta = 3$.}
\end{figure}

We consider next an example of a $(M_1, M_3)$ plane for $m_0$ = 200 TeV, which corresponds to
values of $m_{\tilde q}/m_\chi$ along the plateau in~\cite{ELO}. The left panel of
Fig.~\ref{fig:200} shows the $(M_1, M_3)$ plane for $\tan \beta = 3$ and $A_0/m_0 = 1.5$. In this case there is
a longer gluino coannihilation strip extending nearly all the way to $M_3 \sim 3$~TeV.
The panel on the right again shows the gluino-neutralino mass difference $\Delta M$ (blue line) which in this case peaks
at approximately 170 GeV, which is consistent with the results of \cite{ELO} for intermediate
squark-to-gluino mass ratios.  Also shown is the neutralino mass as a function of $M_3$ (red line):
it again rises to $m_\chi \sim 8$~TeV at the tip of the coannihilation strip, which has $M_3$ slightly $> 3$~TeV.

The Higgs mass in this case is again very slowly varying across the plane and takes the value
$m_H \sim 125$~GeV for this choice of $\tan \beta = 3$. In comparison, had we chosen $\tan \beta = 5$,
the $(M_1, M_3)$ plane would look almost identical but with $m_H \sim 131$~GeV. This and larger values of
$\tan \beta$ are therefore excluded for this value of $A_0/m_0$. We have also studied
smaller values of $A_0/m_0$ and found no consistent solutions of the electroweak
symmetry-breaking conditions for $M_3 \lesssim 500$ GeV for $A_0/m_0 = 1$ and no consistent
solutions across the plane at somewhat lower $A_0/m_0$. This is also the case for $\tan \beta = 10$ and
$A_0/m_0 = 1.5$, for which $m_H \sim 134$~GeV.
Larger values of $A_0/m_0$ also give values that tend to increase $m_H$ and, if increased too much,
the stop becomes the LSP and eventually tachyonic.

\begin{figure}
\begin{center}
\begin{tabular}{c c}
\hspace{-0.6cm}
\includegraphics[height=7cm]{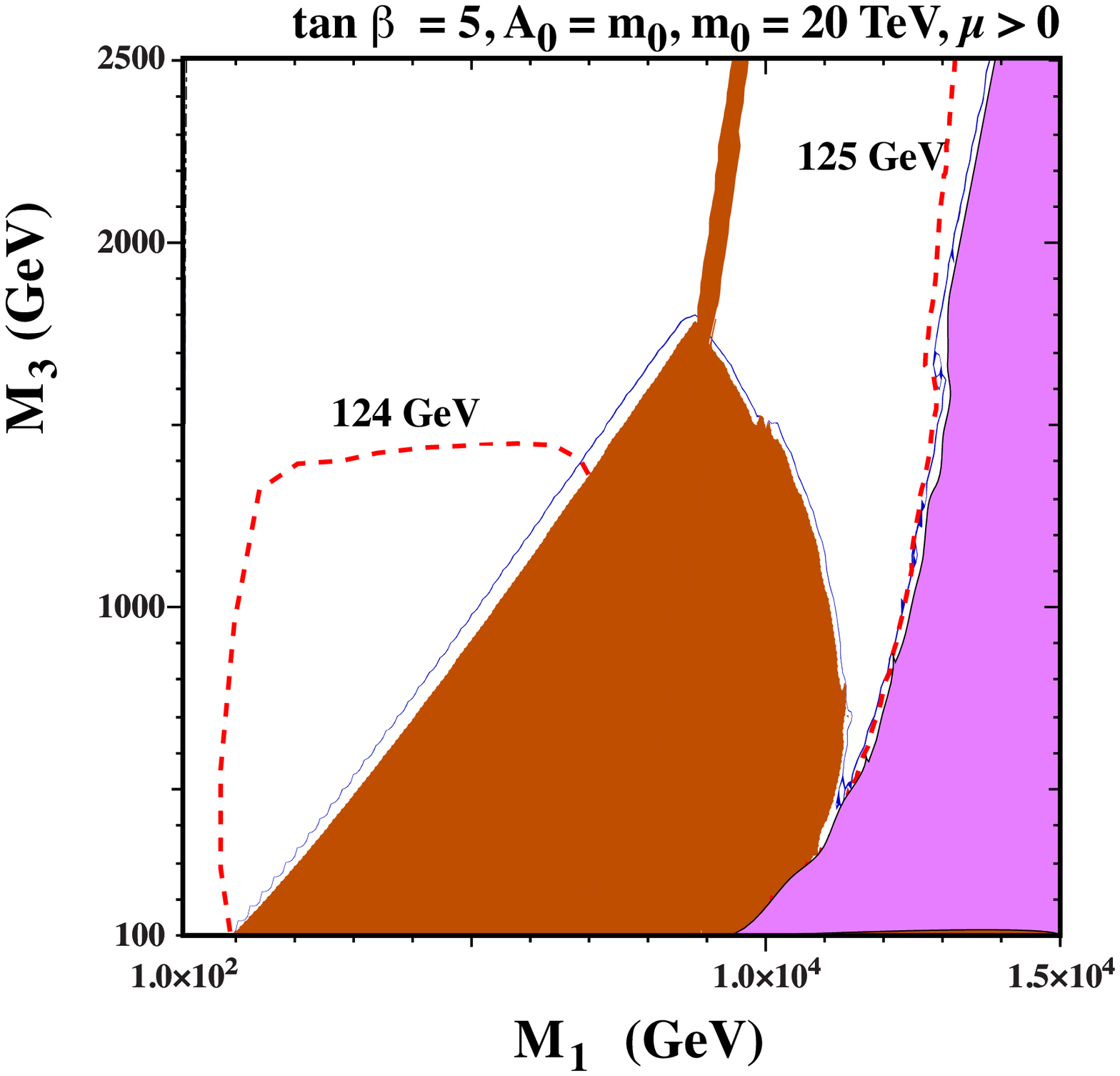} &
\hspace{-0.6cm}
\includegraphics[height=7cm]{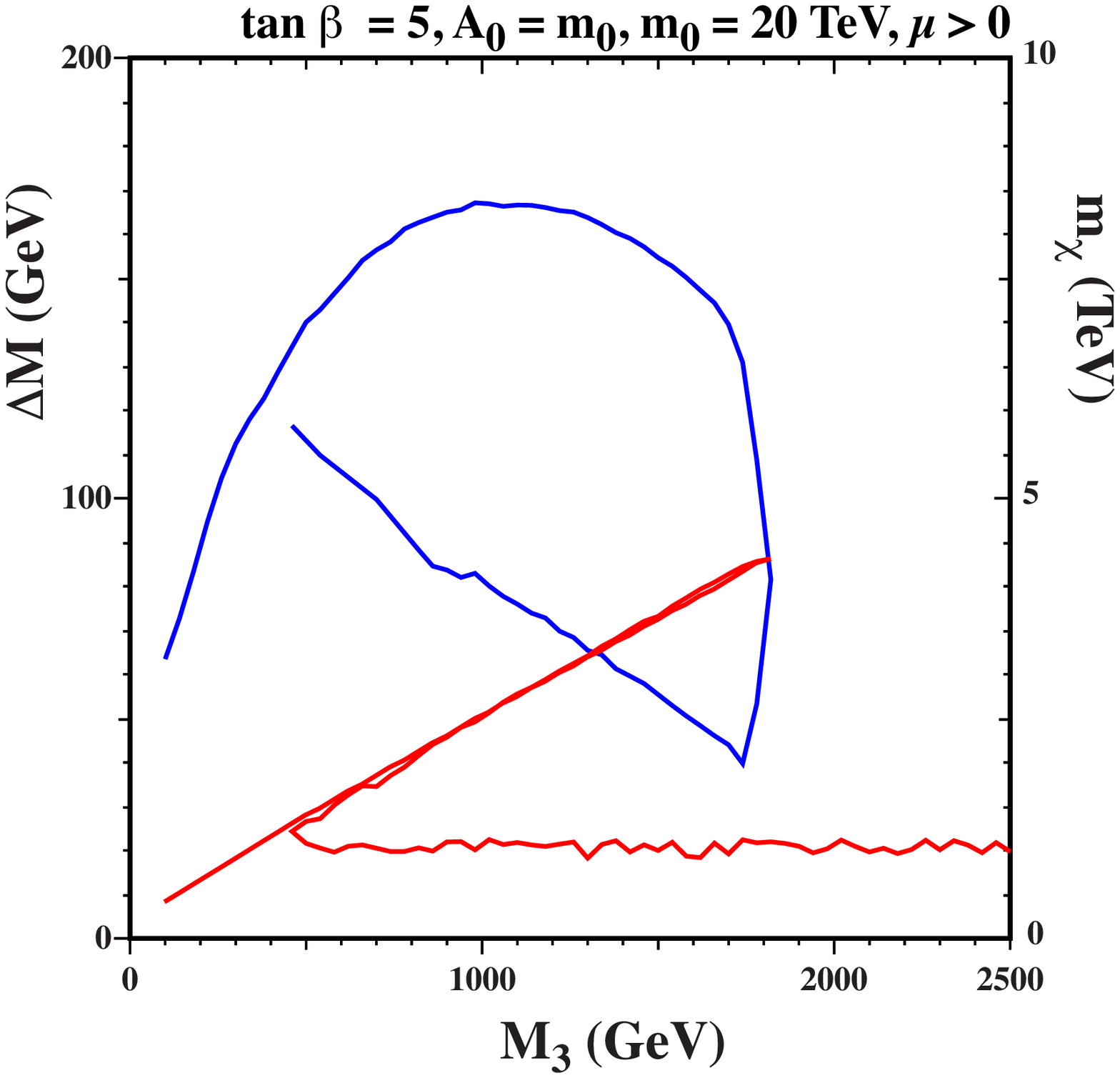} \\
\end{tabular}
\end{center}
\caption{\label{fig:20.2}\it
As for Fig.~\protect\ref{fig:1000}, but for $m_0 = 20$~TeV and $\tan \beta = 5$
with $A_0/m_0 = 1.0$.}
\end{figure}

Next we consider some sample $(M_1, M_3)$ planes with $m_0 = 20$~TeV and $\tan \beta = 5$,
corresponding to the lower end of the $m_\chi$ plateau found in~\cite{ELO}.
Fig.~\ref{fig:20.2} is for the case $A_0/m_0 = 1$, where we
see in the left panel that electroweak symmetry breaking is possible up to values of
$M_1 \lesssim 14$~TeV. There is a gluino coannihilation strip close to the
colored LSP boundary for $M_1 \lesssim 9$~TeV. This is terminated by a spur extending to large
$M_3$ when 9~TeV $\lesssim M_1 \lesssim 10$~TeV, where the lighter
chargino is the LSP.
There is no chargino coannihilation strip along the boundary of this region at large $M_3$, because
the relic density is too high:
for these values of $M_1$ and $M_3$, the Higgsino mass is too large and other coannihilations
are not sufficient to bring the relic density down.
At larger values, 10~TeV $\lesssim M_1 \lesssim$ 11~TeV,
there is a Higgsino-gluino coannihilation strip, which is followed at larger $M_1$ by a
focus-point strip \cite{fp}  hugging the electroweak symmetry breaking boundary where the neutralino
is well-tempered \cite{wt}.
In this case we see both the 124 and 125 GeV Higgs mass contours
and, as in the previous example, $m_H$ is compatible with experiment whenever
the dark matter density falls within the allowed range.

Because the relic density strip is a multi-valued function of $M_3$, the
structure of the gluino-neutralino mass difference $\Delta M$ (blue curve) and the
neutralino mass (red curve) shown in the right panel of Fig.~\ref{fig:20.2} are more complicated than in the previous cases.
After growing to a local maximum $\sim 170$~GeV when $M_3 \sim 1$~TeV, $\Delta M$ starts to fall at larger $M_3$.
We then see a change in behaviour at $M_3 \approx 1800$ GeV along the gluino
coannihilation strip. Here, the neutralino becomes Higgsino-like and, as $M_1$ is increased, the
coannihilation strip tends toward lower $M_3$ with an increasing mass difference,
as seen in the lower branch of the blue curve. A Higgsino LSP emerges for larger $M_1$ because it gives a positive contribution to the up Higgs soft mass from 
renormalization group running. As the up Higgs soft mass goes to zero so does $\mu$ and the Higgsino becomes the LSP. Once $\mu$ is small enough, the Higgsino can be a thermal relic without any assistance in setting the relic density from other particles. In this focus-point-like region, 
the mass difference increases beyond the range displayed.
This behaviour is correlated with the value of $m_\chi$ (red curve),
which increases monotonically to $\sim 4$~TeV. When the bino/Higgsino transition occurs
at $M_3 \approx 1800$ GeV, $m_\chi$ doubles back down to $M_3 \approx 500$ GeV. Then, on the focus-point
branch of the relic density strip, the LSP is mostly Higgsino, the value of
$M_3$ grows, and the lightest neutralino mass takes the
characteristic value $m_\chi \sim 1$ TeV.

\begin{figure}[ht!]
\begin{center}
\begin{tabular}{c c}
\hspace{-0.6cm}
\includegraphics[height=7cm]{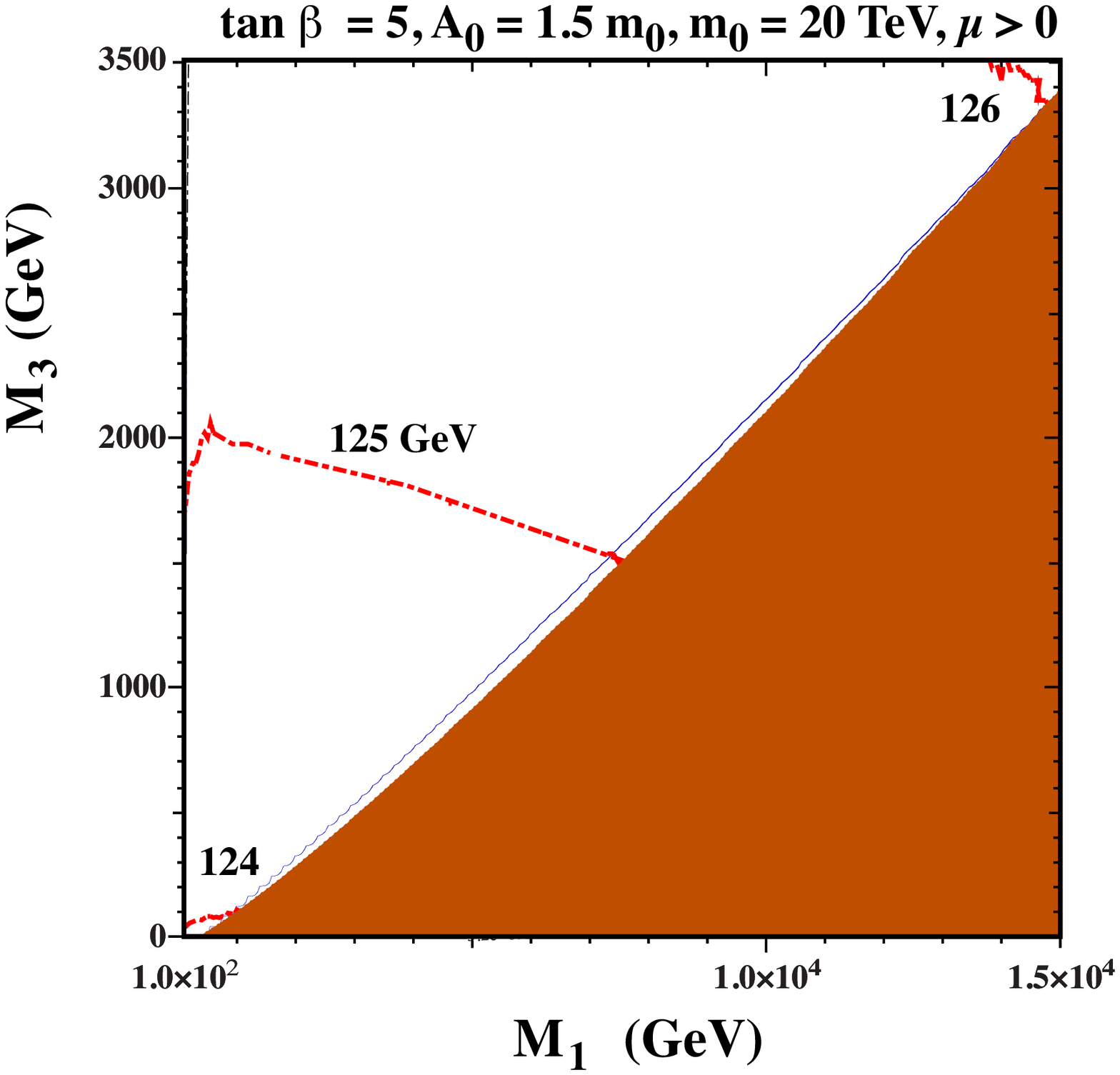} &
\hspace{-0.6cm}
\includegraphics[height=7cm]{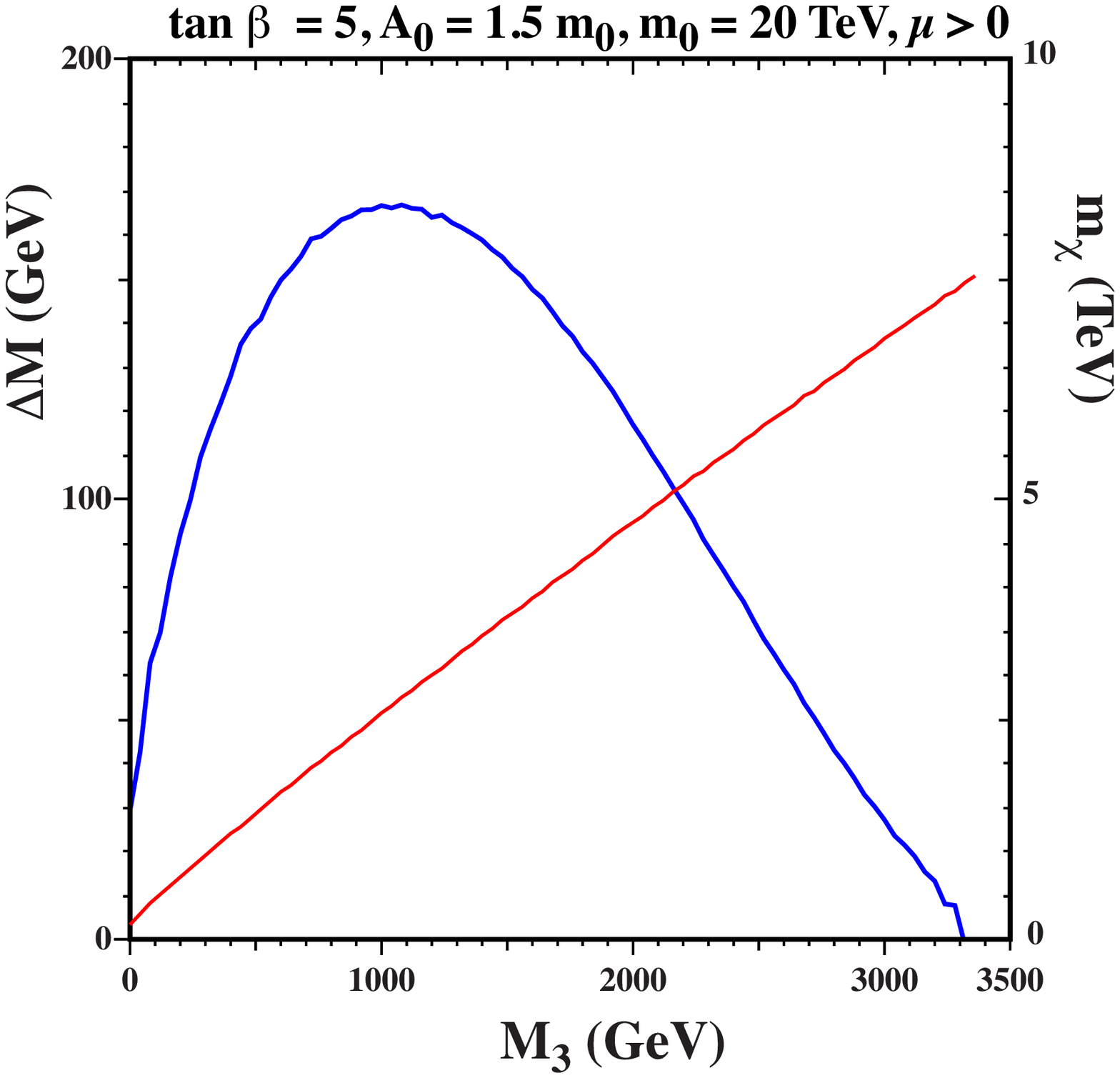} \\

\end{tabular}
\end{center}
\caption{\label{fig:20.3}\it
As for Fig.~\protect\ref{fig:1000}, but for $m_0 = 20$~TeV and $\tan \beta = 5$
with $A_0/m_0 = 1.5$.}
\end{figure}

In Fig. \ref{fig:20.3} we choose a larger value of
$A_0/m_0 = 1.5$, and we see in the left panel a gluino coannihilation strip that extends to
$M_1 \sim 14$~TeV, along which $m_H$ varies between 124 and 126 GeV
as seen by the three Higgs mass contours. The focus-point Higgsino dark matter region has disappeared,
due to the large $A_0$ driving the Higgs mass to large negative values. The end-point of the gluino
coannihilation strip is clearly seen in the right panel of Fig. \ref{fig:20.3},
where $\Delta M \to 0$ (blue curve) at $M_3 \simeq 3300$ GeV.
Qualitatively, this case is similar to that shown in Fig. \ref{fig:200},
rather than to Fig. \ref{fig:20.2} with its truncated gluino coannihilation strip.
In this case, the LSP mass (red curve) rises monotonically to $m_\chi \sim 7.5$~TeV at the end-point of the strip.

\begin{figure}[ht!]
\begin{center}
\begin{tabular}{c c}
\hspace{-0.6cm}
\includegraphics[height=7cm]{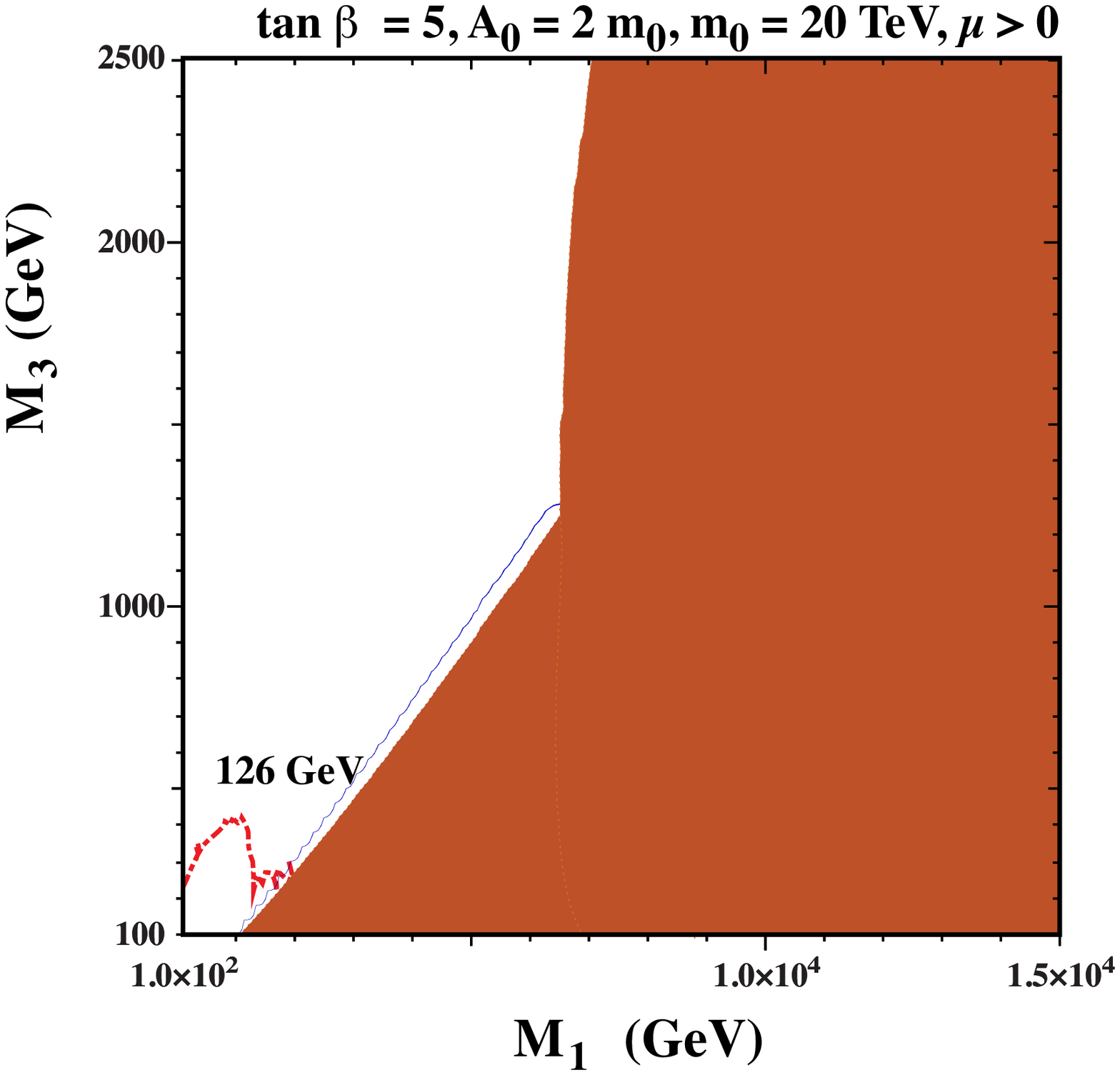} &
\hspace{-0.6cm}
\includegraphics[height=7cm]{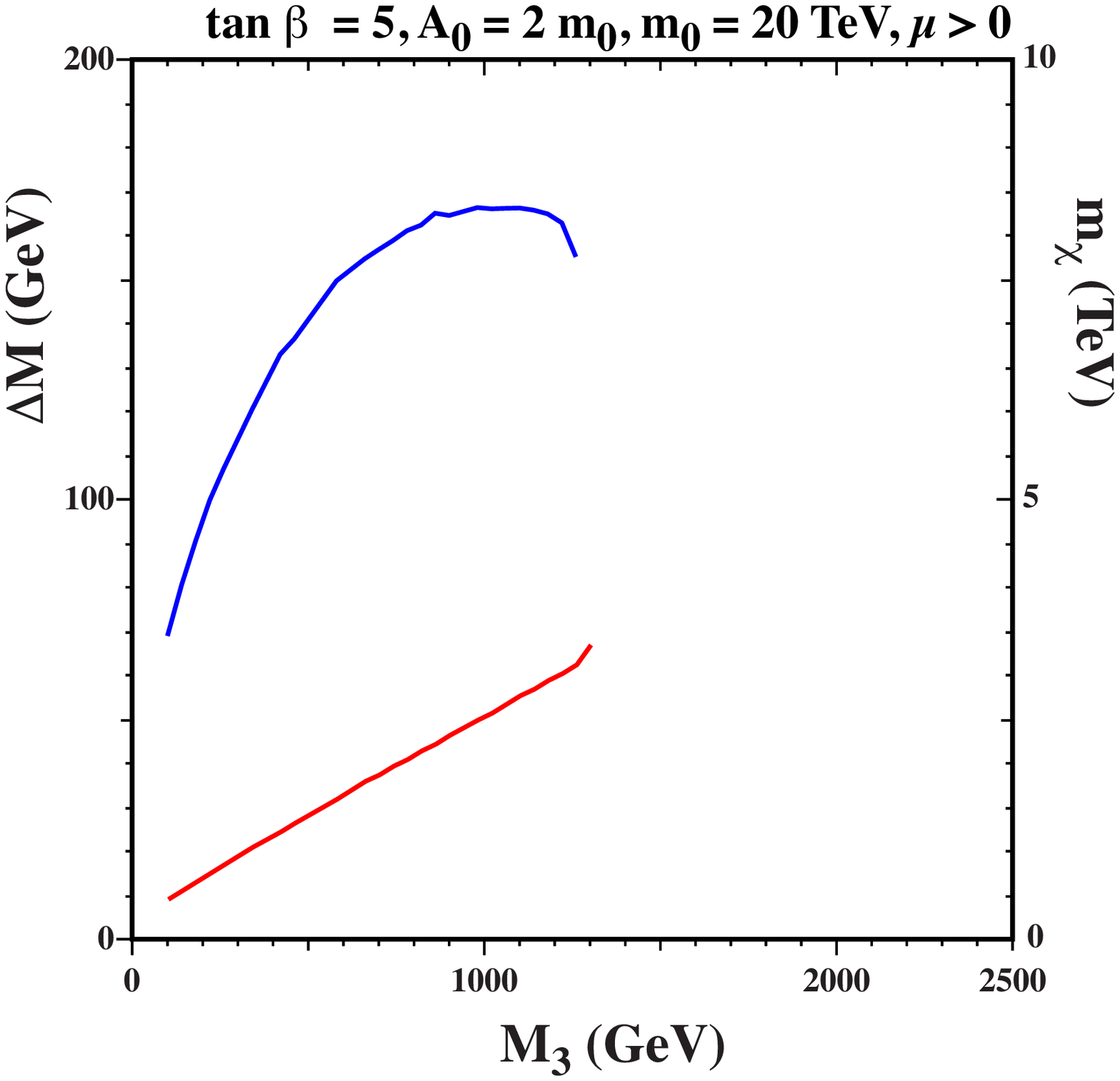} \\

\end{tabular}
\end{center}
\caption{\label{fig:20}\it
As for Fig.~\protect\ref{fig:1000}, but for $m_0 = 20$~TeV and $\tan \beta = 5$
with $A_0/m_0 = 2.0$.}
\end{figure}

In the left panel of Fig.~\ref{fig:20} we
display the $(M_1, M_3)$ plane for $m_0 = 20$~TeV and $A_0/m_0 = 2$. We see again a gluino
coannihilation strip, but extending only to $M_1 \sim 7.5$~TeV. It is terminated
by a stop LSP region that extends to larger values of $M_3$ than those displayed.
In principle, one might have expected to see a stop coannihilation strip
running up along the boundary of the stop LSP region. However, in this case
the relic density is too high along the boundary shown in this figure: as in the chargino case mentioned earlier,
the would-be end-point of the stop coannihilation strip lies within the gluino
LSP region. The value of $m_H$ is generally higher than in the previous case, though
compatible with experiment along all the dark matter strip.
In the right panel of Fig.~\ref{fig:20}, the curves for $\Delta M$ and $m_\chi$ terminate when
the stop becomes the LSP, with $m_\chi \lesssim 3.5$~TeV.

\begin{figure}[ht!]
\begin{center}
\begin{tabular}{c c}
\hspace{-0.6cm}
\includegraphics[height=7cm]{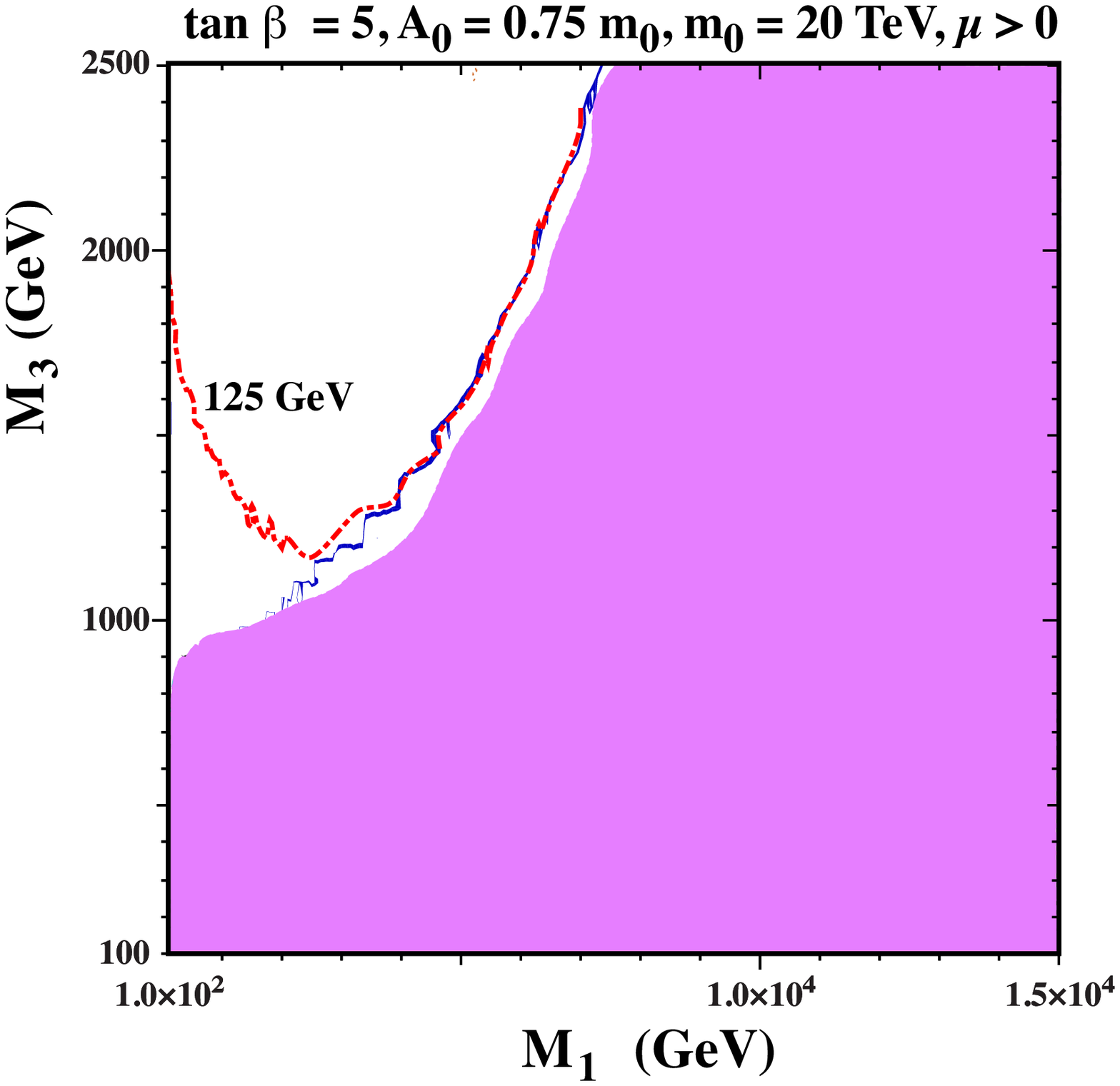} &
\hspace{-0.6cm}
\includegraphics[height=7cm]{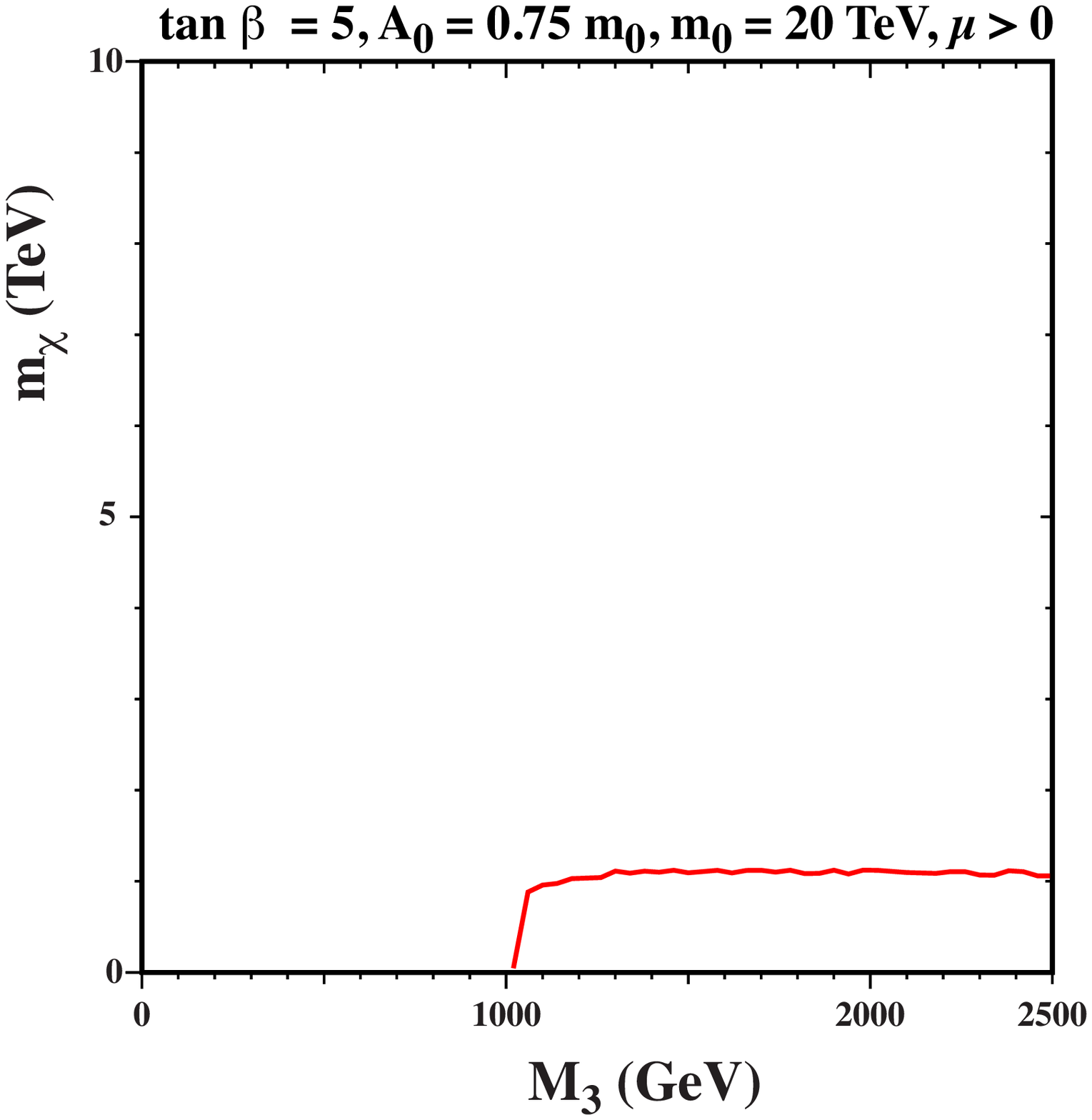} \\
\end{tabular}
\end{center}
\caption{\label{fig:20.1}\it
As for Fig.~\protect\ref{fig:1000}, but for $m_0 = 20$~TeV and $\tan \beta = 5$
with $A_0/m_0 = 0.75$.}
\end{figure}

In Fig.~\ref{fig:20.1}, we choose a lower value of $A_0/m_0 = 0.75$ and keep $\tan \beta = 5$. We see, in the left panel, that consistent electroweak symmetry
breaking is possible only for relatively large $M_3$ and small $M_1$, and
that there is a strip hugging the curved electroweak symmetry-breaking boundary where the LSP has
an enhanced Higgsino component. Its relic density is brought into the
allowed range by the same mechanism as we discussed in the case of a well-tempered neutralino.
As one can see in the right panel of Fig.~\ref{fig:20.1}, once $M_3$ is large enough for EWSB solutions to exist,
the mass of the lightest neutralino (which is mainly a Higgsino) is $m_\chi \sim 1.1$~TeV (red line), almost independent of $M_3$ for
values $\gtrsim 1.1$~TeV. The gluino-neutralino mass difference does not play a role in the relic density
determination and is not shown here.
The red dot-dashed contour shows where $m_H = 125$ GeV: $m_H$ is smaller (larger)
above (below) this contour. The Higgs mass
is highly compatible with the LHC measurement all along the displayed part of the relic density strip.

At larger values of $\tan \beta$, the planes would look similar, though $m_H$ would be larger.
The $(M_1, M_3)$ plane for $m_0 = 20$~TeV
with $\tan \beta = 10$ and $A_0/m_0 = 1$ resembles that in the
left panel of Fig.~\ref{fig:20.1} for $\tan \beta = 5$ and $A_0/m_0 = 0.75$, with a
focus-point strip following closely the curved electroweak symmetry breaking boundary.
The most notable difference is the Higgs mass $m_H$, which is around 128 GeV and only marginally
compatible with experiment after allowing for the theoretical uncertainties. For
the same values of $m_0$ and $\tan \beta =10$,
we find no consistent electroweak symmetry breaking for smaller values of
$A_0/m_0$, and for larger values we find that $m_H$ is too high. Thus we find no
interesting examples of gluino coannihilation for $m_0 = 20$~TeV and $\tan \beta = 10$.

\begin{figure}[ht!]
\begin{center}
\begin{tabular}{c c}
\includegraphics[height=7cm]{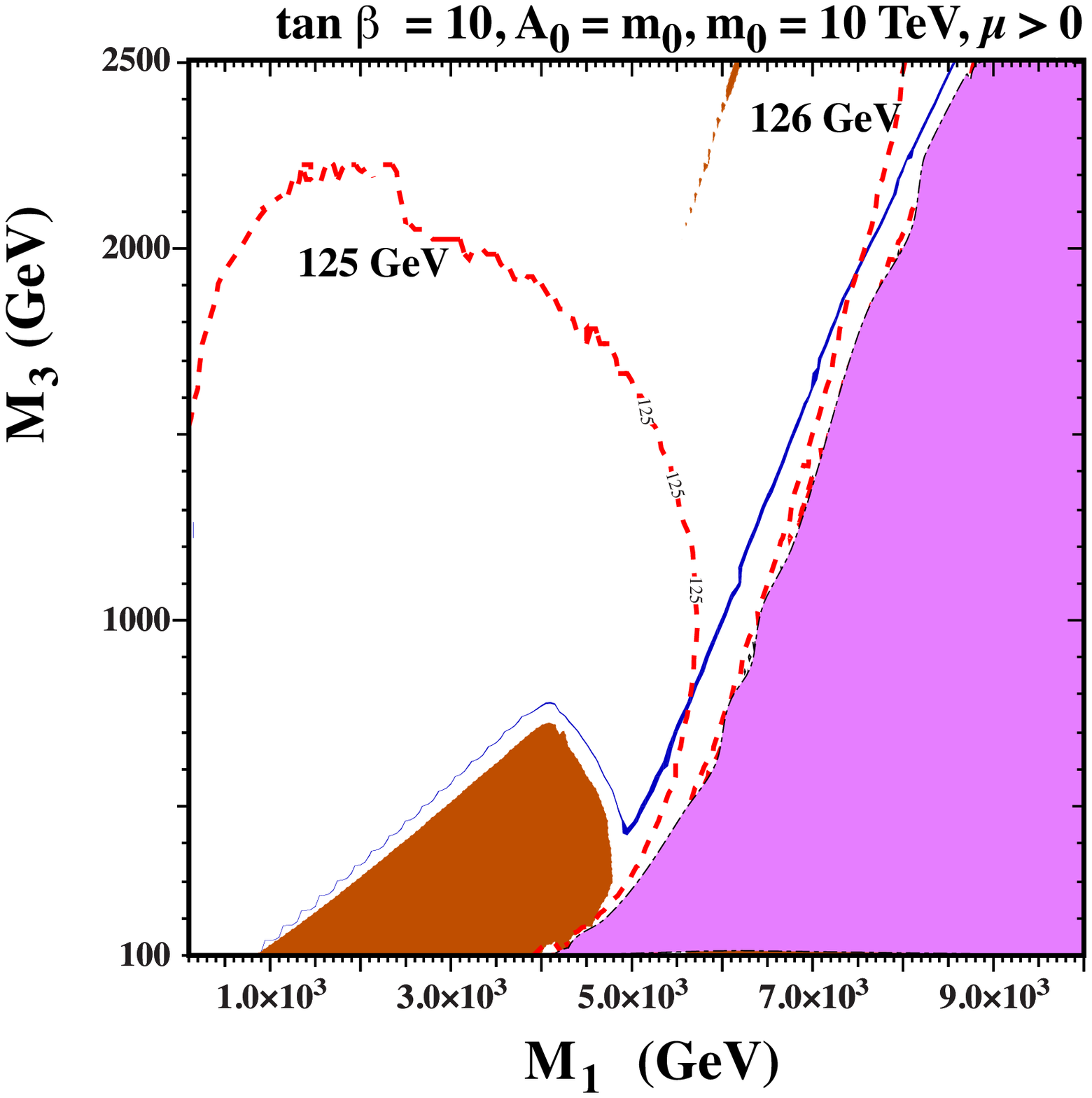} &
\includegraphics[height=7cm]{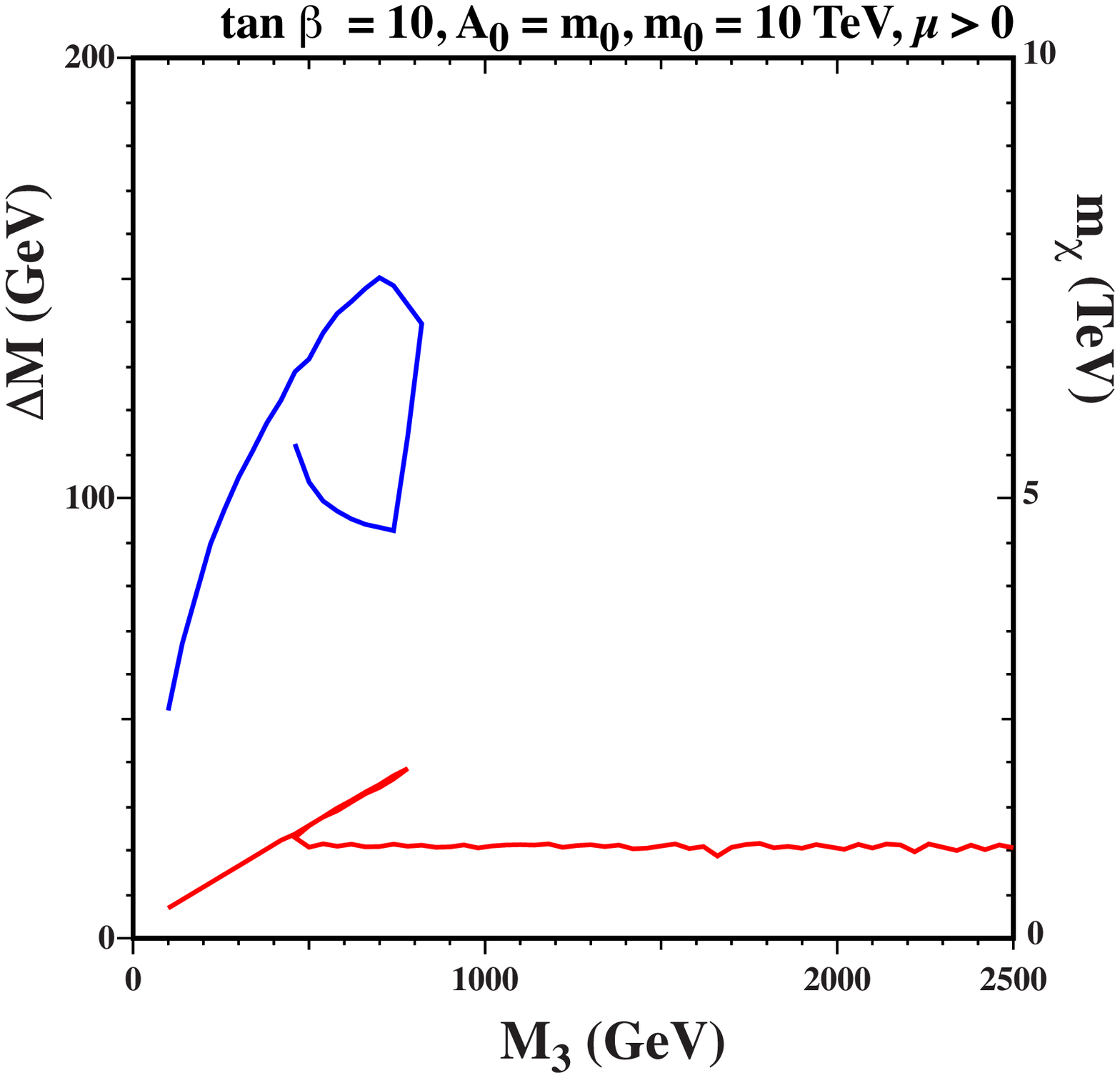} \\
\end{tabular}
\end{center}
\caption{\label{fig:10.1}\it
As for Fig.~\protect\ref{fig:1000}, but for $m_0 = 10$~TeV, $\tan \beta = 10$ and $A_0/m_0 = 1.0$.}
\end{figure}

\begin{figure}[ht!]
\begin{center}
\begin{tabular}{c c}
\includegraphics[height=7cm]{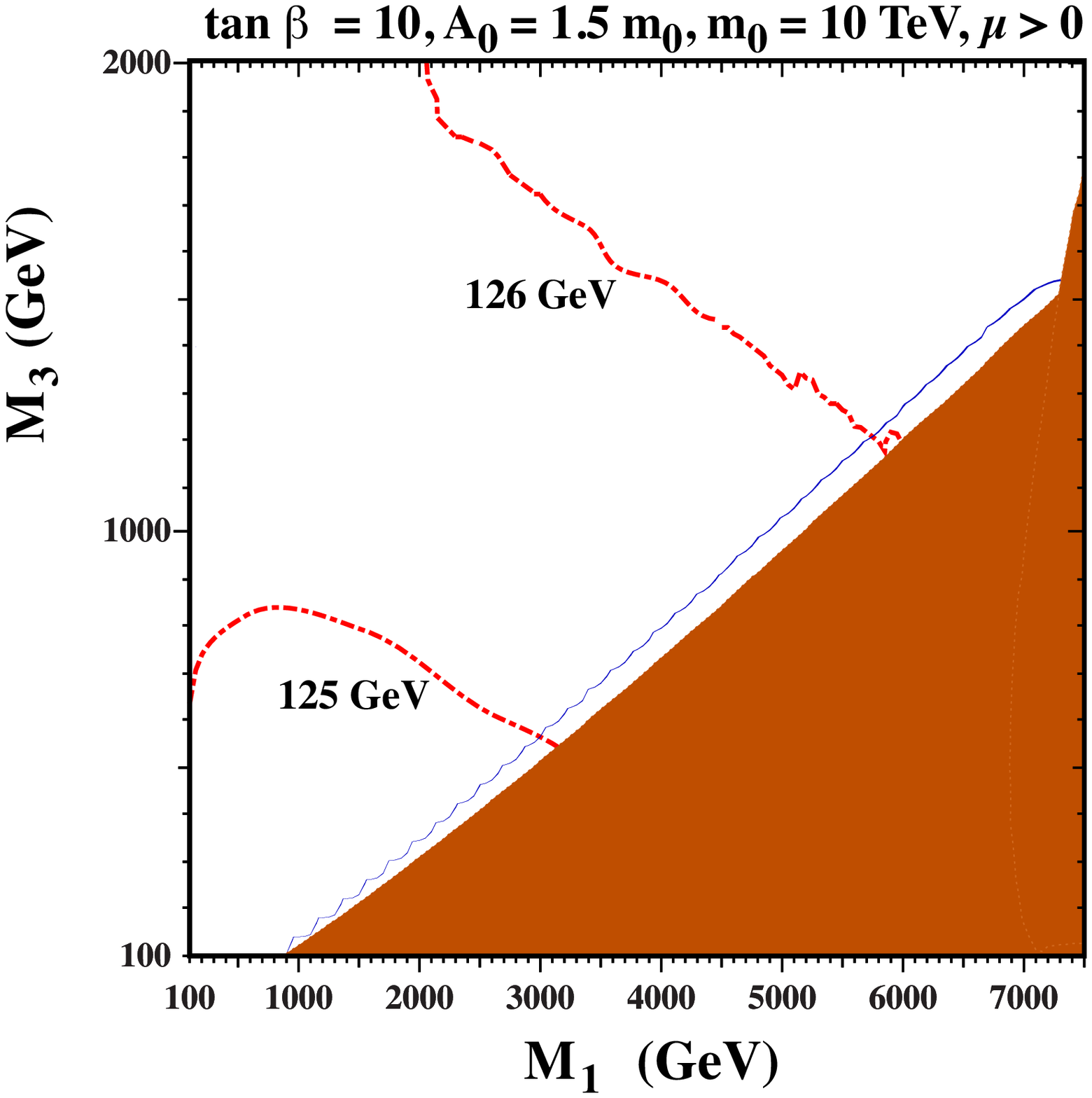} &
\includegraphics[height=7cm]{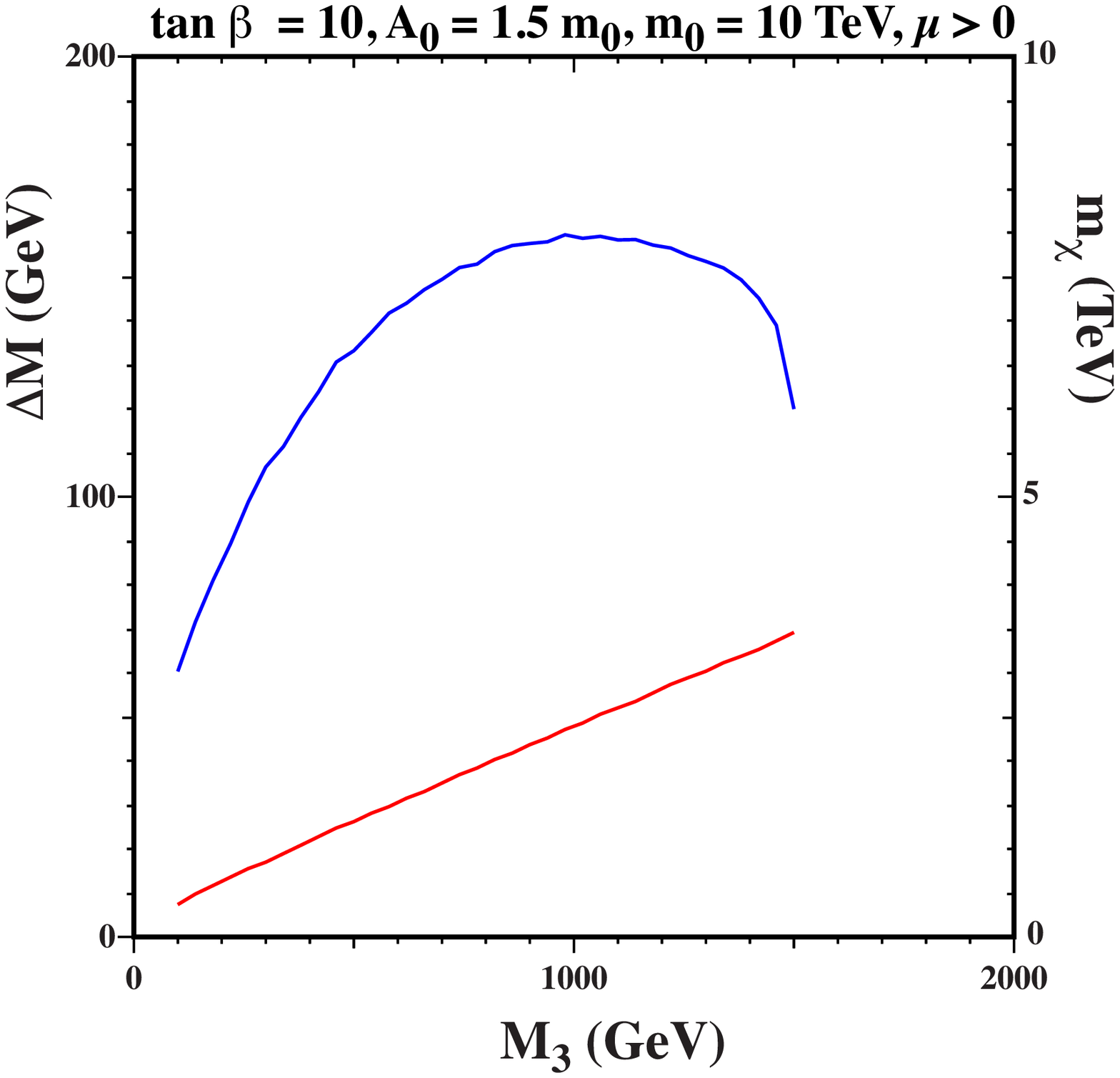} \\
\end{tabular}
\end{center}
\caption{\label{fig:10.2}\it
As for Fig.~\protect\ref{fig:1000}, but for $m_0 = 10$~TeV, $\tan \beta = 10$
and $A_0/m_0 = 1.5$.}
\end{figure}

Finally, we consider in Figs.~\ref{fig:10.1} and \ref{fig:10.2} two examples for $m_0 = 10$~TeV,
corresponding to values of $m_{\tilde q}/m_\chi$ below the $m_\chi$ plateau
in~\cite{ELO}. Fig.~\ref{fig:10.1} is for
$\tan \beta = 10$ and $A_0/m_0 = 1$ and
displays a truncated gluino coannihilation strip extending to $M_1 \sim 4$~TeV,
followed by a Higgsino coannihilation strip extending to $M_1 \sim 5$~TeV, and then a
focus-point strip extending beyond the limits of the plot. We find that $m_H$ is always compatible
with the experimental measurement. This example resembles
that of Fig.~\ref{fig:20.2} for $m_0 = 20$~TeV and $A_0/m_0 = 1$,
the main difference being that the chargino spur has disappeared: we see instead
a chargino LSP island at $M_1 \sim 6$~TeV and $M_3 \gtrsim 2$~TeV.
As in Fig.~\ref{fig:20.2}, we see in the right panel that the gluino-neutralino mass difference (blue curve)
has a multivalued form, and becomes
larger as the neutralino becomes more Higgsino-like along the gluino coannihilation strip.
At larger $M_1$, $M_3$ drops along the Higgsino-gluino coannihilation
strip and the mass difference increases slightly as $M_3$ decreases, increasing beyond the
displayed range as one moves
on to the focus point strip.  Similarly, the neutralino mass rises as $M_1$ is increased, then
 falls back to about 1.1 TeV when the LSP is mostly a Higgsino.

Fig.~\ref{fig:10.2} for $A_0/m_0 = 1.5$ displays a more extended
gluino coannihilation strip reaching $M_1 \sim 7$~TeV and $m_\chi \sim 3.5$~TeV, where it is terminated by
a stop LSP region. This stop LSP region would dominate for larger values of $A_0/m_0$, and the
range of $m_H$ would also become too high.	
At lower $\tan \beta$, the figures would look similar, but with
a smaller Higgs mass. For example, for $\tan \beta = 5$, with $A_0/m_0 = 1.5$ (as in Fig. \ref{fig:10.2}),
the Higgs mass would drop by roughly 3 GeV.
Lower values of $\tan \beta$ would have $m_H$ too small, and lower values of
$A_0/m_0$ but the same value of $\tan \beta$ would have no electroweak symmetry
breaking solutions, while the stop LSP region would dominate for larger $A_0/m_0$.
		
Our analysis of gluino coannihilation in non-universal MSSM
scenarios with $M_1 \ne M_3$ has shown that large values of $m_\chi \lesssim 8$~TeV
are certainly possible,  though restricted by competing mechanisms.
This possibility occurs when $m_0 = 1000$~TeV,
but only for $A_0/m_0 \sim 1.5$ and low values of $\tan \beta \lesssim 3$.  The
possibility of gluino coannihilation becomes more prominent for $m_0 = 200$~TeV,
appearing for an extended range of $\tan \beta$ though still only for $A_0/m_0 \sim 1.5$.
Gluino coannihilation is also prominent for $m_0 = 20$~TeV and $\tan \beta = 5$, but
the focus point may also be important (it becomes dominant when $\tan \beta = 10$),
as is Higgsino coannihilation. These features also appear when $m_0 = 10$~TeV.
In general, this scenario is constrained at small $A_0/m_0$ by the absence of electroweak
symmetry breaking, and at large $A_0/m_0$ by $m_H$ and the appearance of a stop LSP.

\section{Pure Gravity Mediation with Vector Multiplets}

Another possible way of realizing a spectrum that can lead to gluino coannihilation is in models with
pure gravity mediation  \cite{pgm,Hall:2011jd,eioy} of supersymmetry breaking with additional vector multiplets \cite{hari,evo}. The
model we consider here is based on that in \cite{evo}, whose setup we briefly review here.
The effective potential is similar to that for the CMSSM:
 \begin{eqnarray}
V  & =  &  \left|{\partial W \over \partial \phi^i}\right|^2 +
\left( A_0 W^{(3)} + B_0 W^{(2)} + h.c.\right)  + m_{3/2}^2 \phi^i \phi_i^*  \, ,
\label{pot}
\end{eqnarray}
where $W^{(3)}$ corresponds to the trilinear terms of the superpotential,
$W^{(2)}$ contains the bilinear terms of the superpotential, and the $\phi_i$ signify the MSSM fields, with
\begin{eqnarray}
W =  \bigl( y_e H_1 L e^c + y_d H_1 Q d^c + y_u H_2
Q u^c \bigr) +  \mu H_1 H_2  \, .
\label{WMSSM}
\end{eqnarray}
The scalar masses are generated through gravity mediation with a
minimal K\"ahler potential, in an identical manner to mSUGRA \cite{bfs}, and hence are
equal to $m_{3/2}$ at the GUT scale. In general, as seen in (\ref{pot}), the form of $W^{(3)}$ dictates the
pattern of the trilinear supersymmetry-breaking terms. However,
the trilinear couplings are suppressed in PGM models,
because the supersymmetry-breaking field is charged. Thus, we take $A_0=0$.

Since $A_0 = 0$ and the K\"ahler potential is minimal in this model, we have $B_0=-m_{3/2}$ for the Higgs fields.  This leaves three free parameters,
two of which are determined by enforcing the electroweak symmetry-breaking conditions;
\begin{eqnarray}
\mu^2=\frac{m_1^2-m_2^2\tan^2\beta+\frac{1}{2}m_Z^2(1-\tan^2\beta)+\Delta_{\mu}^{(1)}}{\tan^2\beta-1+\Delta_{\mu}^{(2)}} \, ,
\label{eq:mu}
\end{eqnarray}
and
\begin{eqnarray}
B \mu = - \frac{1}{2}(m_1^2+m_2^2+2\mu^2)\sin 2\beta +\Delta_B \, ,
\label{eq:muB}
\end{eqnarray}
where $\Delta_\mu^{(1)}$, $\Delta_\mu^{(2)}$ and $\Delta_B$ are loop corrections to the relationships
\cite{Barger:1993gh}.

This scenario has a very restricted parameter space that does not,
in general, realize electroweak symmetry breaking \cite{eioy}. Therefore, we add
a Giudice-Masiero (GM) term \cite{gm} for the Higgs fields, which modifies the GUT-scale values of both $B_0$ and $\mu$:
\begin{eqnarray}
 \mu &=& \mu_0 + c_H m_{3/2}\ ,
 \label{eq:mu0}
 \\
  B\mu &=&  -\mu_0 m_{3/2} + 2 c_H m_{3/2}^2\ .
   \label{eq:Bmu0}
\end{eqnarray}
This additional degree of freedom in the EWSB sector allows us to choose $\tan\beta$ (in addition to $m_{3/2}$) as a free parameter, and one finds
viable parameter space as long as $\tan\beta\lesssim 3$ \cite{eioy}.

We consider here an extension of this simplest viable version of the PGM scenario
that includes an additional $\mathbf{10}$ and $\mathbf{\overline{10}}$. Because these states are vector-like,
the most general form of the K\"ahler potential will be
\begin{eqnarray}
K=|\mathbf{10}|^2 +|\mathbf{\overline{10}}|^2 + \left(C_{10} (\mathbf{10} \cdot \mathbf{\overline{10}}) +h.c. \right) \, ,
\end{eqnarray}
which includes a GM-like coupling $C_{10}$ that generates a supersymmetric mixing mass
term, $\mu_{10}$, and a supersymmetry-breaking $B$ term for the additional vector-like fields.
Because of the minimal form of the kinetic terms in the limit $C_{10}\to 0$, the additional fields
also have a gravity-mediated tree-level soft supersymmetry-breaking mass equal to $m_{3/2}$.
Since the $\mathbf{{10}}$ contains fields with the same quantum numbers as SM fields,
the $\mathbf{10}$ ($\mathbf{\overline{10}}$) can be combined into gauge-invariant operators with
$H_2$ ($H_1$). If we impose only gauge symmetries the most generic contribution to the superpotential is
\begin{eqnarray}
W=y_t' H_u Q' U' +y_b' H_d\bar Q \bar U  \, ,
\end{eqnarray}
where $Q'$ and $U'$ are from the $\mathbf{10}$ and $\bar Q$ and $\bar U$ are from the $\mathbf{\overline{10}}$.
However, to preserve $R$ symmetry we must take either $y_b'=0$ or $y_t'=0$. Here, we take $y_b'=0$.
The interactions proportional
to $y_t'$ contribute to the beta function of the up Higgs soft mass in a similar way to those controlled by $y_t$.
Specifying a comparable value of $y_t'$ helps drive radiative electroweak symmetry breaking, which
in turn allows larger values of $\tan\beta >3$. The extended theory now has four parameters:
$m_{3/2}$, $\tan \beta$, $C_{10}$, and $y_t'$.

The gaugino masses in the models are generated by anomalies \cite{anom}.
Thus because the contributions to gaugino masses are proportional to $m_{3/2}$,
scalar masses tend to be much heavier than the gaugino masses, reminiscent of split
supersymmetry \cite{split}.  With the addition of the $\mathbf{10}$
and $\mathbf{\overline{10}}$ the anomaly-mediated contributions to the gaugino masses are
\begin{eqnarray}
    M_{1} &=&
    \frac{48}{5} \frac{g_{1}^{2}}{16 \pi^{2}}
    m_{3/2}\ ,
    \label{eq:M1} \\
    M_{2} &=&
    \frac{g_{2}^{2}}{4 \pi^{2}} m_{3/2}  \ ,
        \label{eq:M2}     \\
    M_{3} &=& 0\ .
    \label{eq:M3}
\end{eqnarray}
In addition, the gauginos then get rather large threshold corrections from the $\mathbf{10}$ and
$\mathbf{\overline{10}}$ when they are integrated out, which is in addition to the large threshold correction
coming from integrating out the Higgsinos: for more details see \cite{evo}.
Since the only contribution to the mass of the gluino comes from the threshold corrections,
it tends to be lighter than in typical PGM models. Hence there are regions where the gluino can
coannihilate with the bino, yielding the possibility of a relatively heavy dark matter candidate.

\begin{figure}[ht!]
\begin{center}
\begin{tabular}{c c}
\hspace{-0.6cm}
\includegraphics[height=7cm]{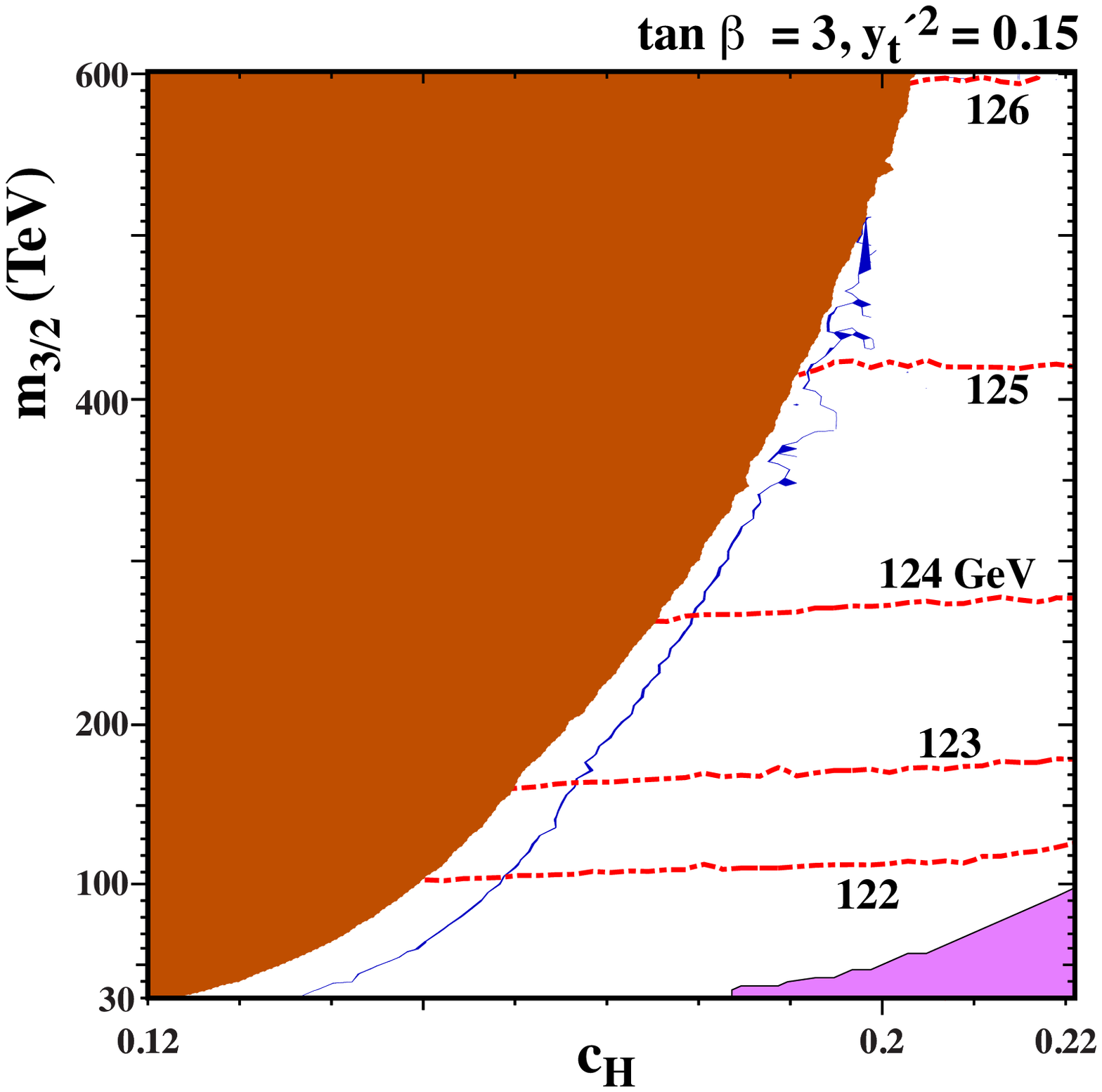} &
\includegraphics[height=7cm]{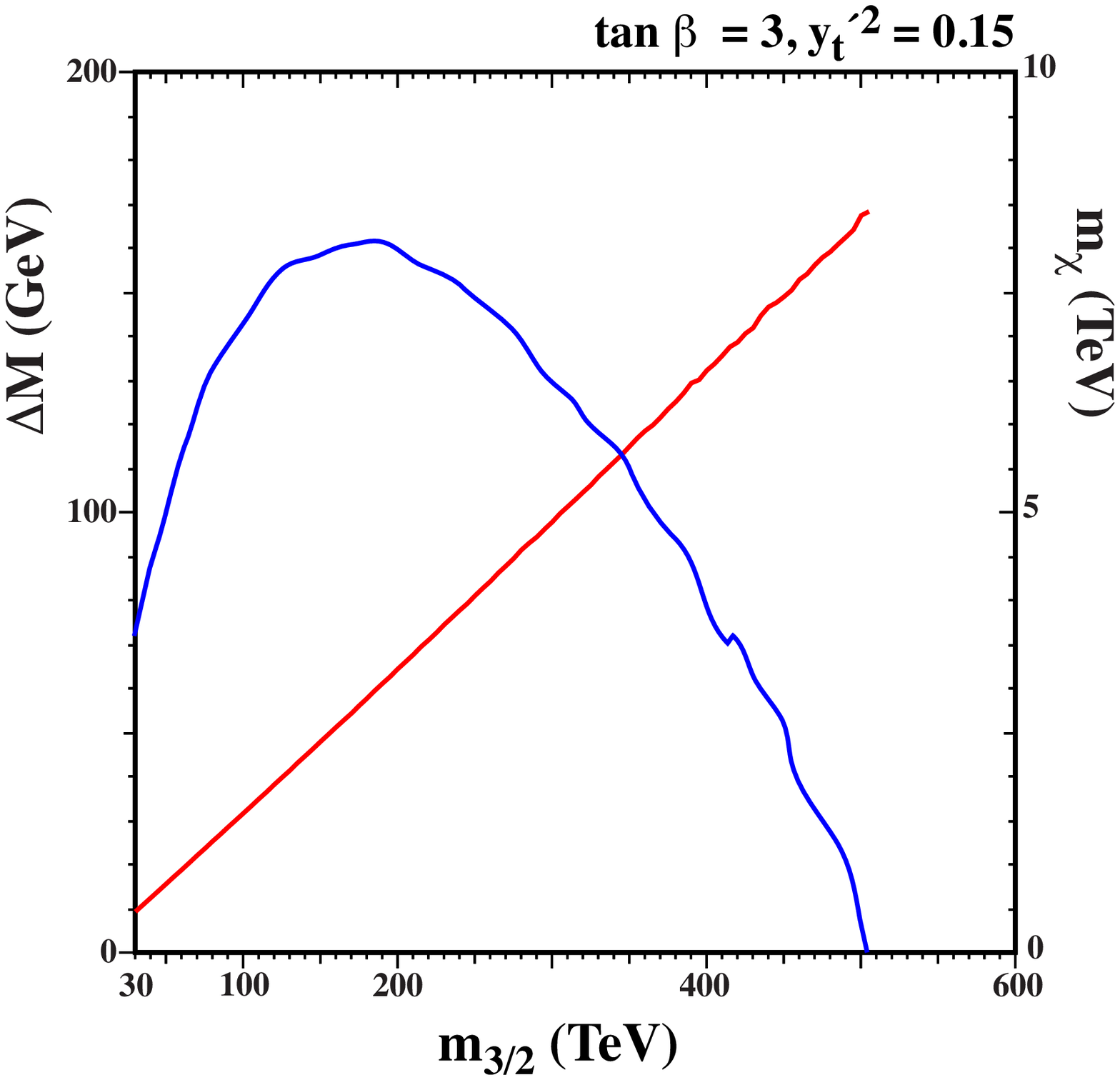}
\vspace{-0.6cm}

\end{tabular}
\end{center}
\caption{\label{fig:EVLSM1}\it
The PGM $(c_H,m_{3/2})$ plane for fixed $\tan \beta = 3$ and $y_t'^2 = 0.15$.
The dark blue strip in the left panel shows where the relic
LSP density $\Omega_\chi h^2$ falls within the $\pm 3$-$\sigma$ range allowed by Planck
and other data, and the lightest neutralino is no longer the LSP in the low-$c_H$ regions shaded brick-red.
One or more of the new vector scalars becomes tachyonic in the lower right corner of the plane (shaded pink).
The right panel shows the gluino-neutralino mass difference (left axis, blue line) and the neutralino mass (right
axis, red line) as functions of $m_{3/2}$.}
\end{figure}

\begin{figure}[ht!]
\begin{center}
\begin{tabular}{c c}
\hspace{-0.6cm}
\includegraphics[height=7cm]{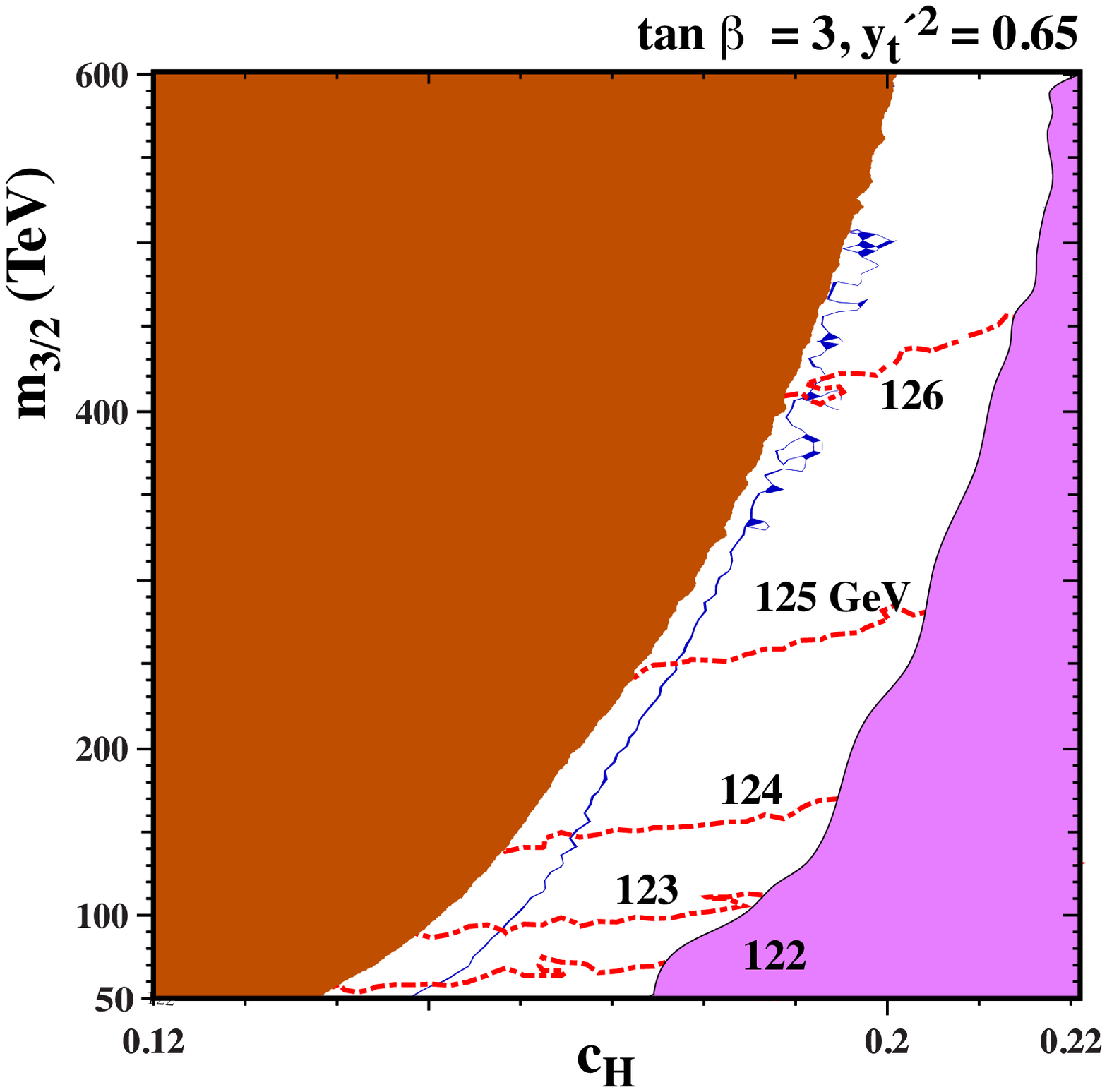} &
\includegraphics[height=7cm]{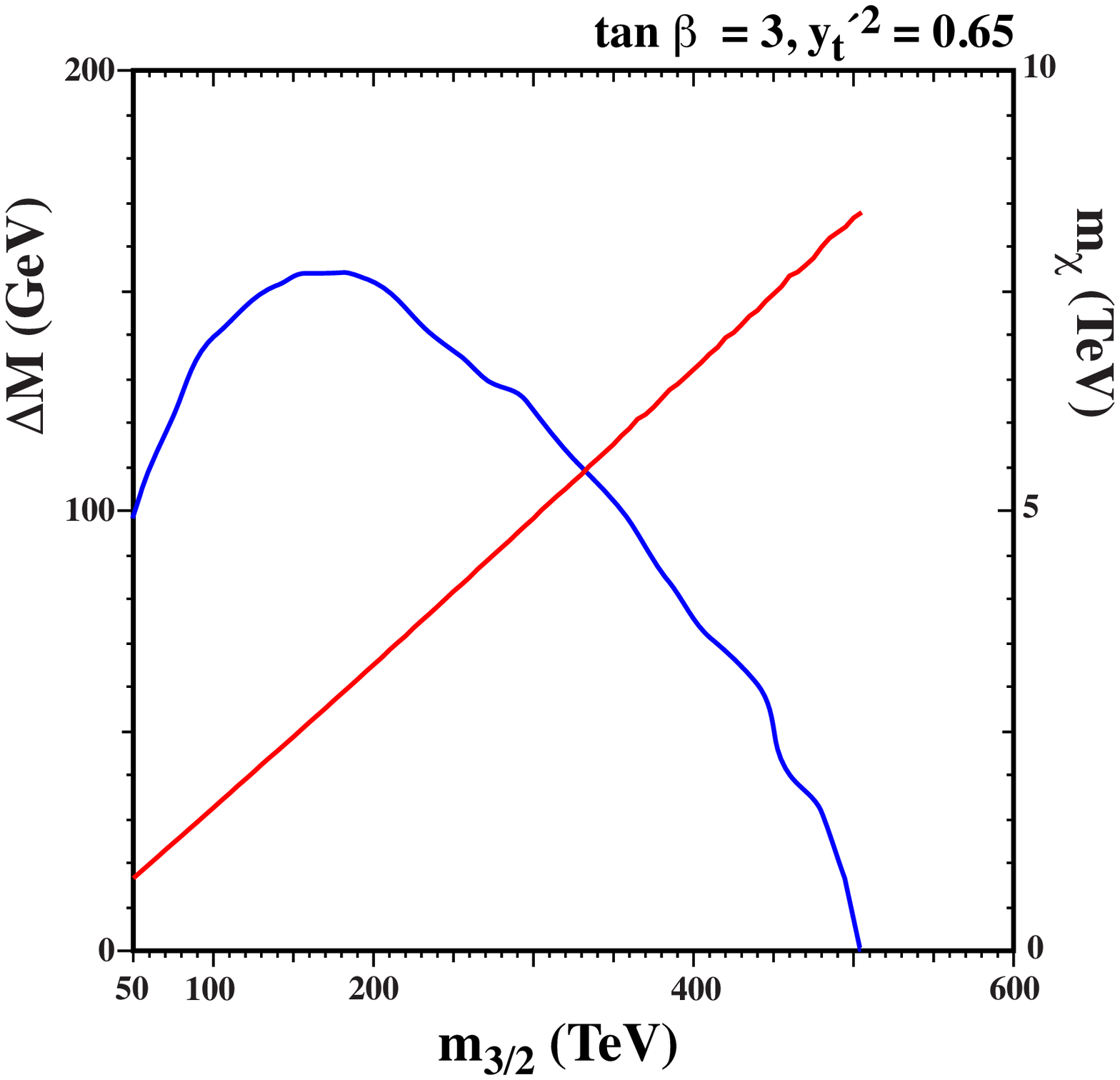}
\vspace{-0.6cm}

\end{tabular}
\end{center}
\caption{\label{fig:EVLSM2}\it
As in Fig. \ref{fig:EVLSM1}, the PGM $(c_H,m_{3/2})$ plane for fixed $\tan \beta = 3$ and $y_t'^2 = 0.65$.
}
\end{figure}

Our results for the PGM model with vector $\mathbf{10}$
and $\mathbf{\overline{10}}$ multiplets can be displayed in $(c_H, m_{3/2})$ planes
for fixed values of the Yukawa coupling, $y_t'$ and $\tan \beta$.  Two examples of these planes are shown in
Figs. \ref{fig:EVLSM1} and \ref{fig:EVLSM2}.
In the former, we have fixed $\tan \beta = 3$ (mainly to get an acceptable value for the Higgs mass, $m_H$,
over the range of $m_{3/2} \le 600$~TeV shown) and $y_t'^2 = 0.15$. In the left panel, we see
a large red shaded region at small $c_H$ where the gluino is the LSP. To the right of this boundary,
we see the gluino coannihilation strip~\footnote{The strip becomes less well defined at
$m_{3/2} \gtrsim 350$ TeV due to inaccuracies of the relic density calculation at such large masses.}.
In the lower right corner, the pink shaded region is excluded
because one or more of the new vector scalars becomes tachyonic. As in previous figures,
the Higgs mass contours are shown as red dot-dashed curves as labelled. Within the
theoretical uncertainties, the Higgs mass agrees with experiment over the part of the plane that is shown.
In the right panel we see, as before, the gluino-neutralino mass difference $\Delta M$ along the gluino
coannihilation strip (blue) and the neutralino mass along the strip (red).
We see that the curve for $\Delta M$ has the same characteristic shape due to strong coannihilations involving the gluino and peaks at $\simeq 170$~GeV  at $m_{3/2} \simeq 200$ TeV when $m_\chi \simeq 3$ TeV.
The end-point of the coannihilation strip occurs at $m_{3/2} \simeq 500$ TeV where $m_\chi \simeq 8.3$ TeV~\footnote{The maximal value
of the LSP mass that is compatible with it being a viable dark matter candidate in this case is similar to the CMSSM case, 
even though there are additional squarks to mediate $\chi_0 q\to q \tilde g$.  This is because the limiting reaction is gluino-gluino
annihilation, which is the same in the two cases.}.

In Fig.~\ref{fig:EVLSM2}, we show in the left panel
corresponding $(c_H,m_{3/2})$ plane for fixed $y_t'^2 = 0.65$ and the same value of  $\tan \beta = 3$.
In this case with a higher Yukawa coupling, slightly higher $c_H$ is needed to obtain a neutralino LSP.
As in the previous case, we see a brick-red shaded gluino LSP for
low $c_H$ and, at slightly larger $c_H$, a gluino coannihilation strip. As
previously, $m_H$ is acceptable along all the displayed portion of the strip
where $m_{3/2} \le 400$~TeV. In this case, the region at larger $c_H$
where one or more of the new vector scalars becomes tachyonic also extends to low $c_H$ for small $m_{3/2}$.
The right panel of Fig.~\ref{fig:EVLSM2} shows the values of $\Delta M$ and $m_\chi$ along the gluino
coannihilation strip. In this case, we see that $\Delta M$ is maximized at $\sim 160$~GeV for $m_{3/2} \sim 150$~TeV. The end-point of the gluino coannihilation strip also occurs at $m_{3/2} \simeq 500$ TeV with
$m_\chi$ around 8.3 TeV.

\begin{figure}[ht!]
\begin{center}
\begin{tabular}{c c}
\hspace{-0.6cm}
\includegraphics[height=7cm]{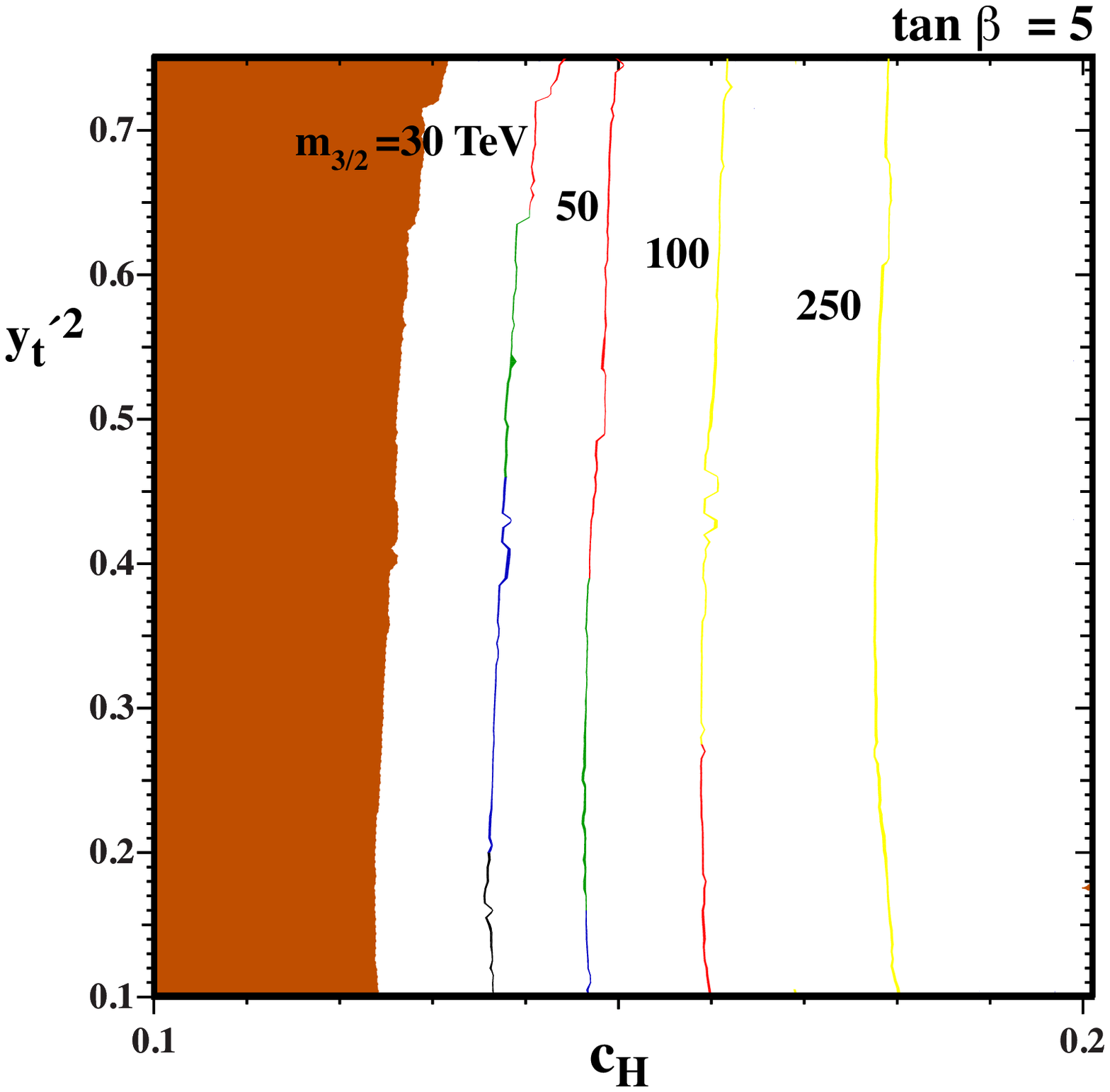} &
\includegraphics[height=7cm]{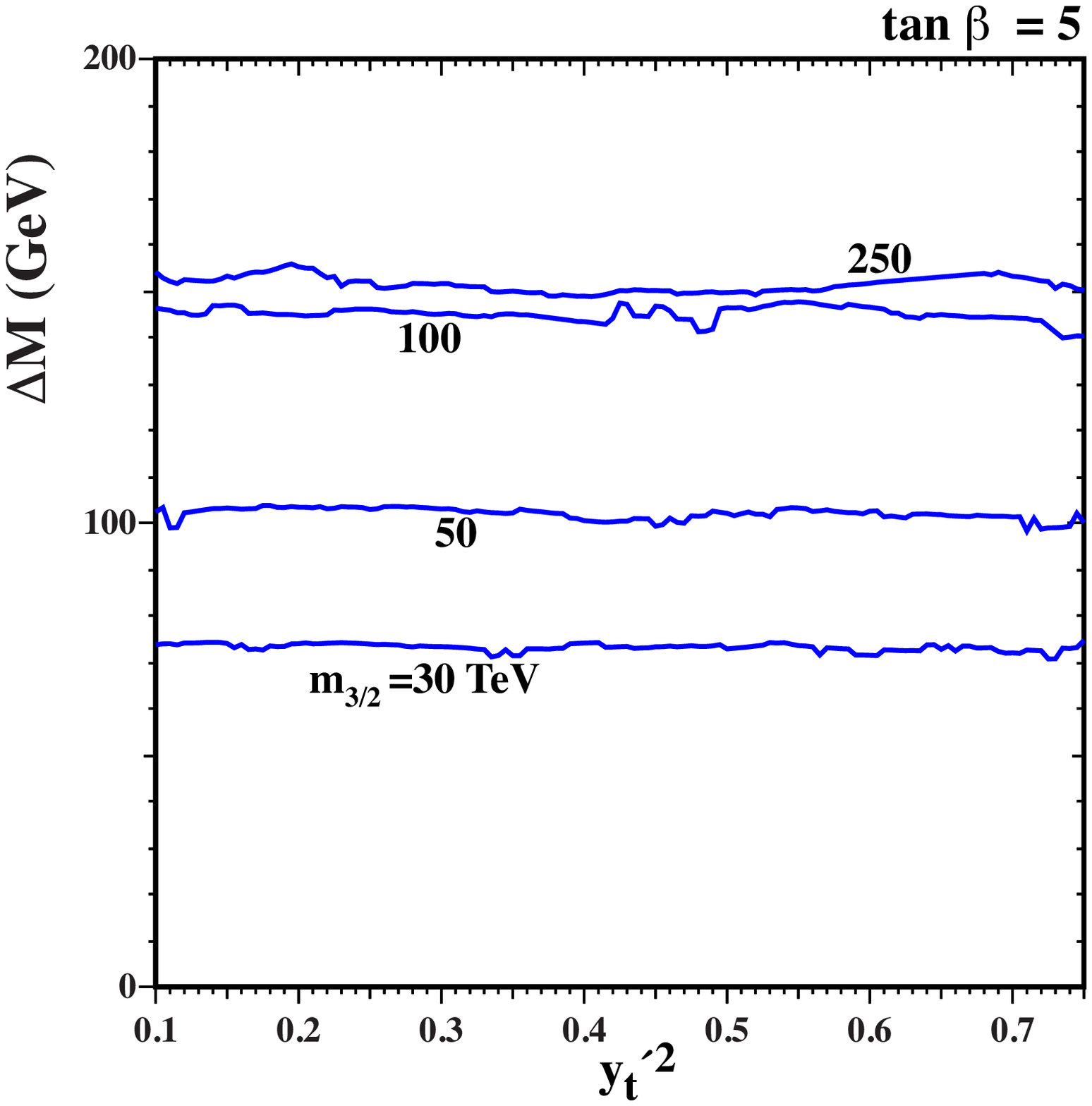}
\vspace{-0.6cm}

\end{tabular}
\end{center}
\caption{\label{fig:EVLSM3}\it
The PGM $(y_t'^2, c_H)$ plane with
$\tan \beta = 5$ for four choices of the gravitino mass, $m_{3/2} = 30, 50, 100$ and 250 TeV.
The right panel shows the gluino-neutralino mass difference as a function of $y_t'^2$.
}
\end{figure}

Finally, we show in Fig. \ref{fig:EVLSM3} an example of a $(y_t'^2, c_H)$ plane with
$\tan \beta = 5$ for four choices of the gravitino mass, namely $m_{3/2} = 30, 50, 100$ and 250 TeV.
The red shaded region has a gluino LSP only in the $m_{3/2} = 30$ TeV case.
In the other three cases this region would be displaced to larger $c_H$. Because the Higgs mass depends on
$m_{3/2}$, there are no unique contours that can be displayed for all four cases.
Instead, we have color-coded the gluino coannihilation strip according to the
Higgs mass: 124-125 GeV (black), 125-126 GeV (blue), 126-127 GeV (green), 127-128 GeV (red),
and $> 128$ GeV (yellow)~\footnote{We recall also that the Higgs mass is sensitive to the choice of $\tan \beta$.}.
The right panel shows that the gluino-neutralino mass difference is almost independent of $y_t'^2$.
We do not show the neutralino mass for these cases, as it is largely independent of $y_t'$ and is
determined from the gravitino mass and can be read from either of the two previous figures.

\section{Summary and Conclusions}

We have presented in this paper a couple of representative MSSM scenarios in which
the gluino may be nearly degenerate with the neutralino LSP $\chi$, whose relic dark matter
density is brought into the range favoured by Planck and other data by gluino
coannihilation. It had been shown previously that values of $m_\chi \lesssim 8~$TeV
are in principle possible when gluino coannihilation is operative~\cite{ELO}, and we have shown in
this paper how such a possibility can be embedded within a scenario for non-universal
soft supersymmetry breaking within the MSSM, on the one hand, and within a simple
extension of the MSSM with pure gravitational mediation of soft supersymmetry
breaking that includes a vector-like $\mathbf{10} + \mathbf{\overline{10}}$
multiplet pair, on the other hand.

In both scenarios, the upper bound on $m_\chi$ depends on the details of the models.
In particular, in the non-universal MSSM scenario there is competition from other mechanisms for
bringing the dark matter density into the Planck range. These, together with other phenomenological
constraints such as the mass of the Higgs boson and the requirement to ensure electroweak
symmetry breaking, restrict the parameter region where gluino coannihilation is dominant.
In the PGM scenario with extra vector-like multiplets, the allowed
range of $m_\chi$ depends on the gravitino mass as well as a vector-like Yukawa coupling.
In both cases, values of $m_\chi \sim 8$~TeV are quite possible.

Gluino coannihilation therefore offers the possibility that the LSP, and hence the
rest of the supersymmetric spectrum, may lie in the multi-TeV range, beyond the reach of the
LHC. Of course, we sincerely hope, if not expect, that supersymmetry will be discovered during
future LHC runs. That said, the scenarios discussed here illustrate one way in which
detection at the LHC could be evaded. An interesting and important question that lies
beyond the scope of this paper is how to detect supersymmetry in a gluino coannihilation
scenario with a multi-TeV LSP (see, e.g., the discussion in~\cite{Nagata:2015hha}). 
As we have discussed in this paper, the gluino-neutralino
mass difference in such a scenario is typically ${\cal O}(100)$~GeV, resulting in a
suppressed missing-energy signature whose detection at a future 100-TeV proton-proton
collider might be challenging. 

{\it Une affaire \`a suivre.}

\section*{Acknowledgements}

The work of J.E. was supported in part by the London Centre for Terauniverse Studies
(LCTS), using funding from the European Research Council via the Advanced Investigator
Grant 267352 and from the UK STFC via the research grant ST/J002798/1. The work of F.L.
was also supported by the European Research Council Advanced Investigator
Grant 267352. The work of J.L.E. and K.A.O.
was supported in part by DOE grant DE-SC0011842 at the University of Minnesota.

\end{document}